\documentclass[letterpaper,titlepage,11pt]{article}
\pdfoutput=1
\usepackage{hyperref}

\usepackage{amssymb,amsmath,amsfonts}
\usepackage{epsfig}
\usepackage{array}
\usepackage{booktabs}

\def\clock{{\count0=\time
          \divide\count0 60
          \ifnum\count0<10 0\fi\the\count0
          \multiply\count0 -60 \advance\count0 \time
          :\ifnum\count0<10 0\fi \the\count0
        }}
\newcommand{\timestamp}{{\small\vbox{\hbox{\tt\jobname.tex}
\hbox{\the\day/\the\month/\the\year, \clock}}}}

\newcommand{\CA}{\mathcal{A}}

\newcommand{\CH}{\mathcal{H}}

\newcommand{\CO}{\mathcal{O}}
\newcommand{\CN}{\mathcal{N}}

\newcommand{\CJ}{\mathcal{J}}

\newcommand{\CV}{\mathcal{V}}

\newcommand{\C}{\mathbb{C}}
\newcommand{\R}{\mathbb{R}}

\newcommand{\gym}{g_{\rm YM}}

\newcommand{\ads}{\mbox{AdS}}

\newcommand{\nn}{\nonumber}

\newcommand{\spa}{\ , \ \ }

\newcommand{\beq}{\begin{equation}}
\newcommand{\eeq}{\end{equation}}

\newcommand{\ds}{\displaystyle}

\newcommand{\tr}{\mathop{{\rm Tr}}}

\newcommand{\matrto}[4]{\left( \begin{array}{cc} #1 & #2 \\
#3 & #4 \end{array} \right) }

\numberwithin{equation}{section}
\begin{document}

\begin{titlepage}
\ \ \vskip 1.8cm
\centerline {\huge \bf Spin Matrix Theory } \vskip 0.4cm
\centerline {\LARGE   A quantum mechanical model of the AdS/CFT correspondence}

\vskip 1.4cm \centerline{\bf Troels Harmark$\,^{1}$ and Marta Orselli$\,^{1,2}$} \vskip 0.7cm

\begin{center}
\sl $^1$ The Niels Bohr Institute, Copenhagen University,  \\
\sl  Blegdamsvej 17, DK-2100 Copenhagen \O , Denmark
\vskip 0.3cm
\sl $^2$ Dipartimento di Fisica, Universit\`a di Perugia,\\
I.N.F.N. Sezione di Perugia,\\
Via Pascoli, I-06123 Perugia, Italy
\end{center}
\vskip 0.3cm

\centerline{\small\tt harmark@nbi.dk, orselli@nbi.dk}

\vskip 1.2cm \centerline{\bf Abstract} \vskip 0.2cm \noindent
We introduce a new quantum mechanical theory called Spin Matrix theory (SMT). The theory is interacting with a single coupling constant $g$ and is based on a Hilbert space of harmonic oscillators with a spin index taking values in a Lie (super)algebra representation as well as matrix indices for the adjoint representation of $U(N)$. We show that SMT describes $\CN=4$ super-Yang-Mills theory (SYM) near zero-temperature critical points in the grand canonical phase diagram. Equivalently, SMT arises from non-relativistic limits of $\CN=4$ SYM. Even though SMT is a non-relativistic quantum mechanical theory it contains a variety of phases mimicking the AdS/CFT correspondence. Moreover, the $g \rightarrow \infty$ limit of SMT can be mapped to the supersymmetric sector of string theory on $\ads_5\times S^5$. We study $SU(2)$ SMT in detail. At large $N$ and low temperatures it is a theory of spin chains that for small $g$ resembles planar gauge theory and for large $g$ a non-relativistic string theory. When raising the temperature a partial deconfinement transition occurs due to finite-$N$ effects. For sufficiently high temperatures the partially deconfined phase has a classical regime. We find a matrix model description of this regime at any coupling $g$. Setting $g=0$ it is a theory of $N^2+1$ harmonic oscillators while for large $g$ it becomes $2N$ harmonic oscillators.

%\vskip 0.5cm \leftline{\timestamp}
\end{titlepage}

%\pagestyle{empty}
%\small
%\begin{spacing}{1}
%\small
\tableofcontents
%\end{spacing}
%\tableofcontents
%\normalsize
%\newpage
\pagestyle{plain}
\setcounter{page}{1}

%%%%%%%%%%%%%%%%%%%%%%%%%%%%%%%%%%%%%%%%%%%%%%%%%%%%%%%%%%
\section{Introduction and summary}
\label{sec:intro}

The AdS/CFT correspondence between $\CN=4$ super-Yang-Mills theory (SYM) with gauge group $SU(N)$ and type IIB string theory on $\ads_5\times S^5$ promises in its strongest version a complete quantitative agreement between the two theories for any $N$ and any 't Hooft coupling $\lambda$ \cite{Maldacena:1997re}. %,Gubser:1998bc,Witten:1998qj}.
Recent years of research have improved enormously our quantitative understanding of the AdS/CFT correspondence in two sectors of the theory. One is the supersymmetric sector with the technique of localization that enables one to compute exact partition functions \cite{Pestun:2007rz}. Another is the planar limit with $N=\infty$ for which one employs an integrable spin chain as the connecting link between weakly coupled planar $\CN=4$ SYM and tree-level string theory on $\ads_5\times S^5$ \cite{Minahan:2002ve,Beisert:2003tq,Beisert:2006ez}. In that case it is the presence of the integrability symmetry that enables one to quantitatively interpolate between the two sides of the correspondence.

The goal of this paper is to devise a way to go beyond these two sectors in order to obtain a quantitative understanding of the AdS/CFT correspondence with $N < \infty$, both for supersymmetric and non-supersymmetric observables, enabling one to interpolate between weak and strong 't Hooft coupling. The motivations for this are many. An important one is that to study black holes in AdS/CFT one needs to go beyond infinite $N$, and include non-perturbative effects in $1/N$ for large $N$, $e.g.$ what one can call finite-$N$ effects. Understanding black holes quantitatively in the AdS/CFT correspondence would be of enormous importance, particularly if one can go beyond the supersymmetric sector. Similarly, to study the emergence of D-branes in the AdS/CFT correspondence, for example in the form of Giant Gravitons, one needs to understand finite-$N$ effects as well.
 
The idea of this paper is to generalize the integrable spin chain as connecting link between the gauge and string theory sides beyond $N=\infty$. However, since it appears that the integrability symmetry does not in general extend beyond $N=\infty$ one needs another simplifying feature to enable one to realize this idea.%
\footnote{There is evidence of integrability symmetry for excitations of Giant Gravitons \cite{Carlson:2011hy}. However, it is not clear that one can make a general extension of the integrability symmetry from $N=\infty$ to large $N$. Indeed there are indications that the symmetry breaks down for $1/N$ corrections \cite{Beisert:2003tq,Kristjansen:2010kg}.} The simplifying feature will be to consider the AdS/CFT correspondence in certain non-relativistic limits that in the grand canonical ensemble correspond to approaching critical points at zero temperature $T=0$. Let $\vec{\Omega}$ parametrize the five chemical potentials conjugate to the relevant global symmetry charges of $\CN=4$ SYM, then we take a limit of the form \cite{Harmark:2006di,Harmark:2006ta,Harmark:2006ie,Harmark:2007px,Harmark:2008gm}
\begin{equation}
\label{introlimit}
(T , \vec{\Omega}) \rightarrow (0,\vec{\Omega}^{(c)}) \spa \lambda \rightarrow 0 \spa \mbox{with} \ \frac{\lambda}{T} \ \mbox{and}\ \frac{\vec{\Omega}-\vec{\Omega}^{(c)}}{T} \ \mbox{kept fixed}
\end{equation}
where $\vec{\Omega}^{(c)}$ parametrizes the critical point. For $N=\infty$ the result of this limit is that one gets a much simpler spin chain as connecting link that only has a nearest neighbor interaction and which is a non-relativistic quantum mechanical theory. We also get a rescaled coupling $g$ proportional to the 't Hooft coupling such that for small $g$ the spin chain resembles planar $\CN=4$ SYM in a subsector and for large $g$ a non-relativistic string theory which can be obtained as a limit of string theory on $\ads_5\times S^5$ \cite{Kruczenski:2003gt,Harmark:2008gm}. Most importantly, the spin chain theory is so simple that it is possible to take the strong coupling limit $g \gg 1$ without need of employing the integrability symmetry.

The central proposal of this paper is that Spin Matrix theory provides the connecting link between the gauge and string theory sides in the AdS/CFT correspondence near a zero-temperature critical point. Spin Matrix theory is a new non-relativistic quantum mechanical theory that we define in this paper. It can be thought of as a finite-$N$ generalization of nearest-neighbor spin chains. 
Spin Matrix theory is based on a Hilbert space of harmonic oscillators with both a spin index and a matrix index. The matrix index belongs to the adjoint representation of $U(N)$.%
\footnote{For simplicity we base Spin Matrix theory on the $U(N)$ group rather than $SU(N)$. See Section \ref{sec:discuss} for a translation of our results to $SU(N)$.} Instead the spin index is in a semi-simple Lie (super)algebra representation and for $N=\infty$ Spin Matrix theory reduces to a nearest-neighbor spin chain based on this representation. It includes an interacting Hamiltonian with a single coupling constant $g$. 

\begin{figure}[h!]
\centerline{\includegraphics[scale=0.43]{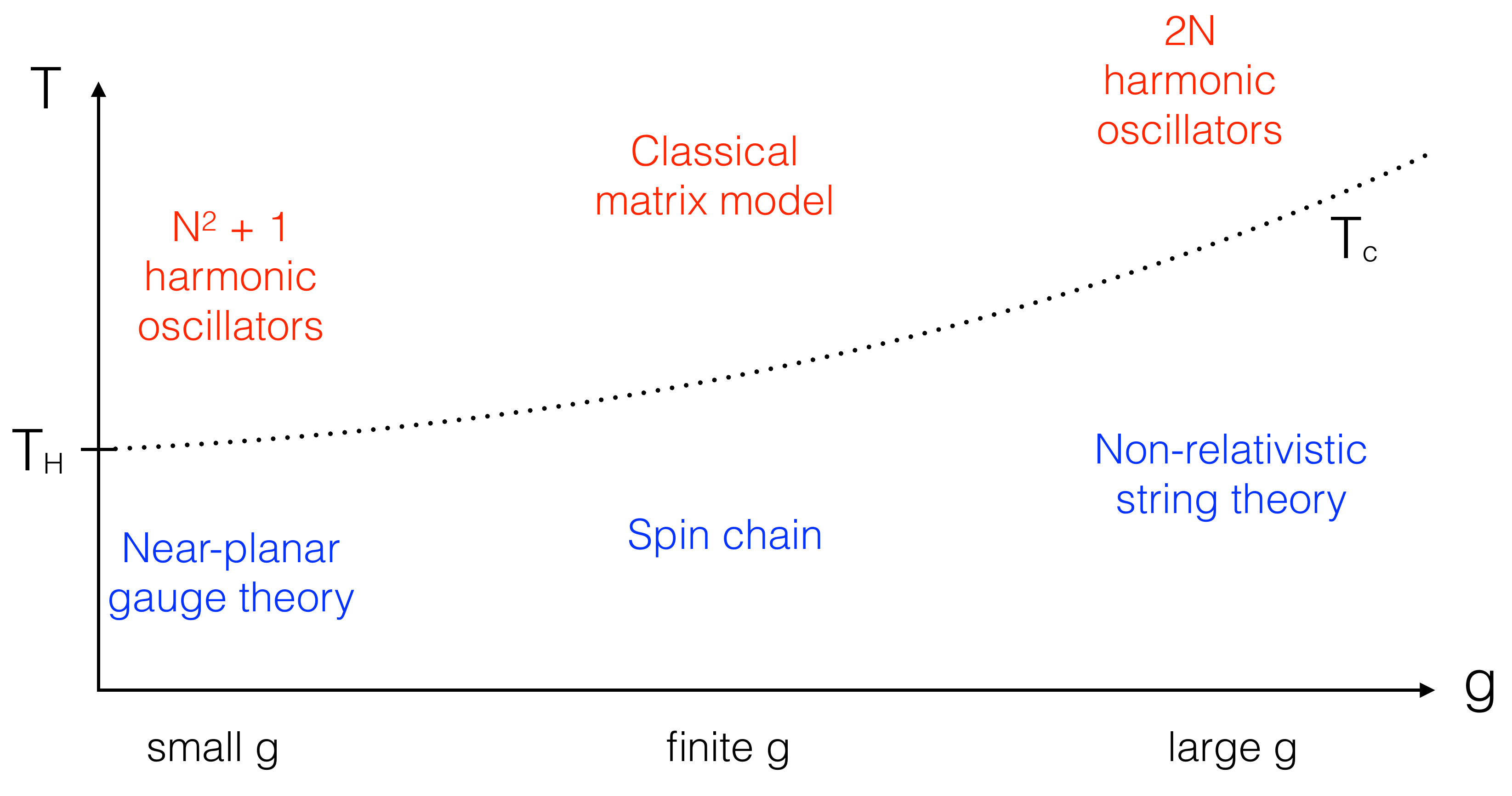}}
\caption{\small Phase diagram of $SU(2)$ Spin Matrix theory as function of the temperature $T$ and the coupling $g$. The stipled line marks the temperature $T_c$ where a partial deconfinement transition occurs. At zero coupling this meets the Hagedorn temperature $T_H$.}
\label{fig:SMT}
\end{figure}

We show that $\CN=4$ SYM with gauge group $U(N)$ near zero-temperature critical points, in the sense of the limit \eqref{introlimit}, indeed is described by particular versions of Spin Matrix theory for any given $N$. 
For a particular zero-temperature critical point we get what we denote as $SU(2)$ Spin Matrix theory.

We study in detail the phase diagram of the $SU(2)$ Spin Matrix theory in this paper. 
Despite that it is a non-relativistic quantum mechanical theory with a relatively simple formulation it includes a variety of very interesting phases. Taking a fixed but large $N$ one can parametrize the phase diagram in terms of the temperature $T$ and the coupling constant $g$. We have illustrated the phase diagram in Figure \ref{fig:SMT}. For small temperatures and any $g$ $SU(2)$ Spin Matrix theory is described as a gas of (weakly interacting) Heisenberg spin chains. For small $g$ this can be described as near-planar $\CN=4$ SYM in the $SU(2)$ sector at weak coupling, while for large $g$ it can be described as a non-relativistic string theory (with small string coupling). Note in particular that the semi-classical limit of the non-relativistic string theory is accurately described at tree-level by the Landau-Lifshitz sigma-model. As explained in \cite{Harmark:2008gm} this is no coincidence as the limit \eqref{introlimit} for the $SU(2)$ critical point can be reinterpreted as the limit of Kruzcenski \cite{Kruczenski:2003gt} of string theory on $\ads_5\times S^5$.

Raising the temperature $T$ of the weakly interacting spin chain gas the perturbative $1/N$ effects give rise to an increasing interaction among the spin chains. Eventually finite-$N$ effects come into play and for sufficiently high temperatures one encounters a partial deconfinement transition at a temperature which we denote $T_c$, as illustrated by the stipled line in Figure~\ref{fig:SMT}.
We explore the partially deconfined phase above this phase transition by considering its behavior in the high-temperature regime where we find that its description simplifies. 
For zero coupling $g=0$ we analyze in detail the partition function of $SU(2)$ Spin Matrix theory and find that for sufficiently large temperatures it reduces to the partition function of $N^2+1$ uncoupled harmonic oscillators. This happens in the classical limit where we can view the harmonic oscillators as distinguishable.  

Turning on the coupling $g$ we show using coherent states that $SU(2)$ Spin Matrix theory at sufficiently high temperatures is described by a classical matrix model. This classical matrix model is based on four Hermitian $N\times N$ matrices in the Hamiltonian formulation. For $g=0$ it describes $N^2+1$ uncoupled harmonic oscillators as mentioned above. For non-zero $g$ the matrix model has a potential term proportional to $g$ that gives rise to interactions between the $N^2 +1$ harmonic oscillators. At large $g$ most of the harmonic oscillators become infinitely heavy and decouple, leaving a phase of $2N$ uncoupled harmonic oscillators. 

We see $SU(2)$ Spin Matrix theory as a quantum mechanical model for the AdS/CFT correspondence since its phases bear strong resemblence to phases of the AdS/CFT correspondence. For low temperature the Heisenberg spin chain works as a connecting link between small and large coupling, as already mentioned, with clear connections to the small and large 't Hooft coupling limits of the AdS/CFT correspondence at $N=\infty$. For high temperatures we see a partial deconfinement into a high-temperature phase described by a classical matrix model for interacting harmonic oscillators. Our assertion is that this resembles a phase corresponding to a highly excited gas of D-branes.

\begin{figure}[h!]
\centerline{\includegraphics[scale=0.6]{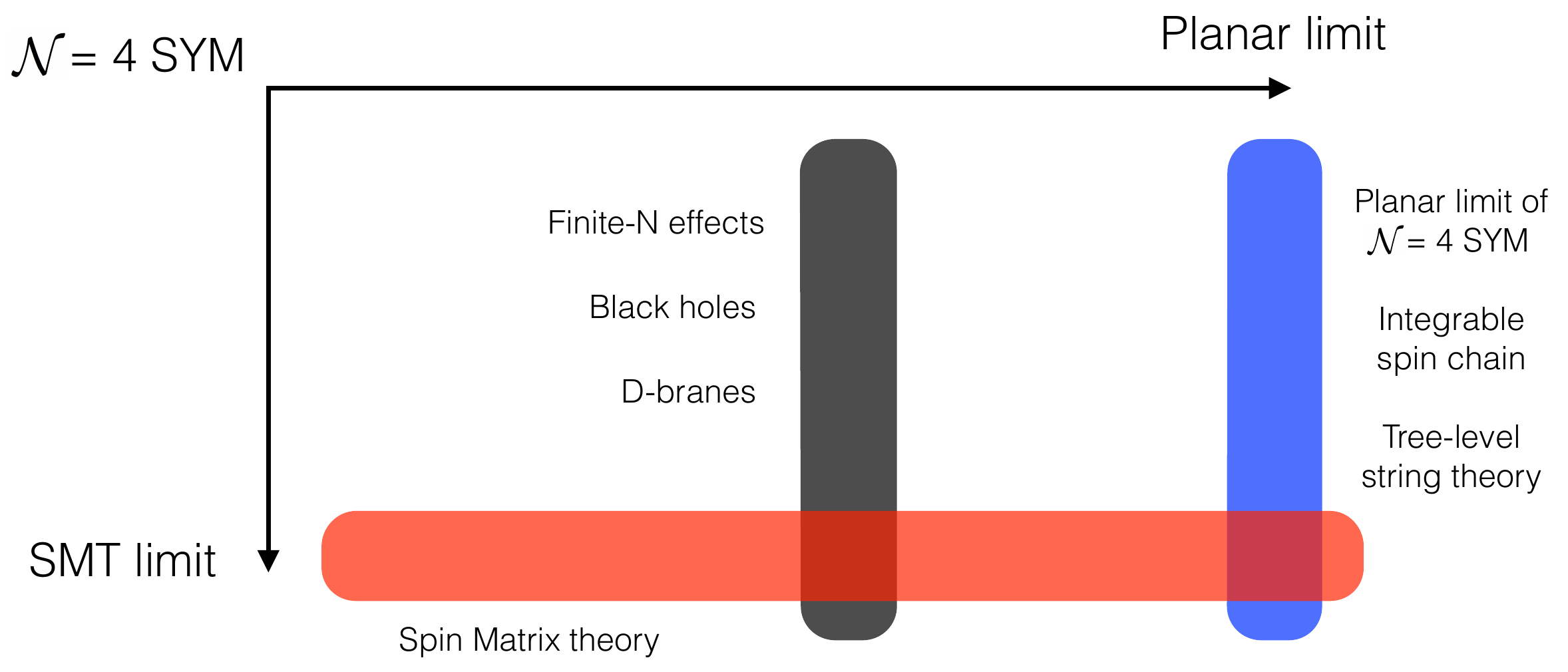}}
\caption{\small Illustration of our general philosophy for Spin Matrix theory. The diagram represents the regimes of $\CN=4$ SYM. Towards the right one approaches the planar limit regime, depicted in blue. Towards the bottom one approaches the Spin Matrix theory regime depicted in red. The black area depicts the regime in which one finds black holes and D-branes.}
\label{fig:philosophy}
\end{figure}

We have summarized our general philosophy of how we envision that the Spin Matrix theory limits of $\CN=4$ SYM can be used to approach a quantitative understanding of finite-$N$ effects in the context of the AdS/CFT correspondence in Figure \ref{fig:philosophy}. The regime with black holes and D-branes is separated from the planar limit regime in the figure since it is non-perturbative in the string coupling, and hence corresponds to finite-$N$ effects, whereas in the planar limit the string coupling is exactly zero (hence also Newtons constant is exactly zero). Instead the Spin Matrix theory limit gives an effective rescaled string coupling $g/N$ that can be tuned, thus making it possible to have overlaps of the Spin Matrix theory regime with the two other regimes.%
\footnote{In the related context of the gauge/gravity duality for D0-branes impressive work has been done to make numerical simulations on the gauge theory side that approximately reproduce the gravity side (see for instance the recent work \cite{Hanada:2013rga}). While this gives important evidence for holography one would ultimately like to have an analytical approach.}  

In Section \ref{sec:g_infty} we confirm this philosophy in the supersymmetric sector by showing for a zero-temperature critical point of $\CN=4$ SYM that one can match the $g\rightarrow \infty$ limit of Spin Matrix theory exactly to what one obtains for string theory on $\ads_5\times S^5$ at the dual critical point (assuming the validity of a commonly accepted conjecture). In Section \ref{sec:discuss} we elaborate more on this philosophy, and argue that it can be employed also beyond the supersymmetric sector.

The structure of this paper is as follows. In Section \ref{sec:SMT} we define Spin Matrix theory in general and show that it has a limit as nearest neighbor spin chain theory for $N=\infty$. In Section \ref{sec:SMTlimit} we show that Spin Matrix theory describes $\CN=4$ SYM near critical points, or, equivalently, can be obtained from $\CN=4$ SYM in non-relativistic limits. In Section \ref{sec:lowT} we focus on $SU(2)$ Spin Matrix theory and review results in the planar and near-planar limits corresponding to low temperature. In Section \ref{sec:hightemp} we derive new results on the high temperature behavior of $SU(2)$ Spin Matrix theory where finite-$N$ effects sets in. In Section \ref{sec:discuss} we discuss our results.

%%%%%%%%%%%%%%%%%%%%%%%%%%%%%%%%%%%%%%%%%%%%%%%%%%%%%%%%%%
\section{Spin Matrix theory}
\label{sec:SMT}

\subsection{Definition of Spin Matrix theory}
\label{sec:defSMT}

Spin Matrix theory is a quantum mechanical theory with a
well-defined Hilbert space and Hamiltonian acting on the Hilbert
space. Spin Matrix theory is built on a representation $R_s$ of a semi-simple Lie (super-)group $G_s$, which we here call the {\sl spin group}, and on the adjoint representation $R_m$ of the group $U(N)$ on the space of $N \times N$ complex matrices.%
\footnote{One can generalize this to matrix representations of other groups such as $SU(N)$, $SO(N)$ and $Osp(N)$.}

\subsubsection*{Hilbert space of Spin Matrix theory}

We consider first the purely bosonic case. Define the raising operators
\begin{equation}
\label{raisop}
{(a^\dagger_{s})^i}_j
\end{equation}
Here $s \in R_s$ is in the representation of the spin group $G_s$ and the $i,j$ indices are $N \times N$ indices corresponding to the adjoint representation $R_m$  of $U(N)$ ($i=1,...,N$ labels the fundamental and $j=1,...,N$ the anti-fundamental representation of $U(N)$). Corresponding to the raising operators \eqref{raisop} we have the vacuum $|0 \rangle$ and the lowering operators ${(a^{s})^j}_i$ such that
\begin{equation}
\label{bosrel}
{(a^{s})^j}_i | 0 \rangle = 0 \spa \Big[ {(a^{s})^j}_i , {(a^\dagger_{s'})^k}_l \Big] = \delta^s_{s'} \delta_{i}^k \delta^{j}_l
\end{equation}
and all raising operators commute with each other. This defines a bosonic harmonic oscillator for each $s\in R_s$ and $(i,j)\in R_m$. Hence we can define a Hilbert space $\CH'$ as all the possible harmonic oscillator states 
\begin{equation}
\CH' = \sum_{L=1}^\infty \mbox{sym}  \Big[ (R_s \otimes R_m)^L \Big]
\end{equation}
involving the symmetric product of $L$ representations $R_s \otimes
R_m$.
We can write a basis for $\CH'$ as
\begin{equation}
\label{thebasisprime}
{(a^\dagger_{s_1})^{i_1}}_{j_1} {(a^\dagger_{s_2})^{i_2}}_{j_2} \cdots {(a^\dagger_{s_L})^{i_L}}_{j_L} | 0 \rangle \spa L=1,2,...
\end{equation}
The Hilbert space $\CH$ of Spin Matrix theory is defined as the linear subspace of $\CH'$ of states that are singlets of the $R_m$ representation. The singlet condition on a state $|\phi \rangle$ in $\CH'$ is
\begin{equation}
\label{singletcond}
\Phi^i{}_j |\phi \rangle = 0 \spa \Phi^i{}_j \equiv \sum_{s \in R_s} \sum_{k=1}^N \Big[ (a^\dagger_s)^i {}_k (a^s )^k{}_j - (a^\dagger_s)^k{}_j (a^s)^i{}_k \Big]
\end{equation}
One finds that the Hilbert space $\CH$ is spanned by the set of states of the form
\begin{equation}
\label{thebasis} \sum_{i_1,i_2,...,i_L = 1}^N
{(a^\dagger_{s_1})^{i_1}}_{i_{\sigma(1)}} {(a^\dagger_{s_2})^{i_2}}_{i_{\sigma(2)}} \cdots {(a^\dagger_{s_L})^{i_L}}_{i_{\sigma(L)}} | 0 \rangle   \spa L=1,2,...
\end{equation}
where $\sigma \in S(L)$ is an element of the permutation group
$S(L)$ of $L$ elements. Using a slightly different notation we can equivalently say $\CH$ is spanned by the set of states
\begin{equation}
\label{tracebasis}
\tr ( a^\dagger_{s_1} a^\dagger_{s_2} \cdots a^\dagger_{s_{l}} ) \tr ( a^\dagger_{s_{l+1}} \cdots )
 \cdots \tr ( a^\dagger_{s_{k+1}} \cdots a^\dagger_{s_L} ) | 0 \rangle \spa L = 1,2,...
\end{equation}
where the traces are over the $R_m$ indices. The individual cycles of the permutation elements correspond to single traces. 
In general one can find linear relations between the states of the form \eqref{thebasis} or \eqref{tracebasis} when $L > N$. To have a proper basis for the Hilbert space one would need to thin out the set of states such that only a linearly independent set is left. Such a basis is provided by the restricted Schur polynomials which in addition are orthogonal \cite{Balasubramanian:2004nb,Koch:2007uu}. 

One can also include fermionic excitations. This is realized as a split up $R_s = B_s \oplus F_s$ of the spin group representation. Then for $s \in B_s$ the raising operator ${(a^\dagger_{s})^i}_j$ and the corresponding lowering operator obey the bosonic commutator \eqref{bosrel}. Instead for $s \in F_s$ we have
\begin{equation}
\label{fermrel}
{(a^{s})^j}_i | 0 \rangle = 0 \spa \Big\{ {(a^{s})^j}_i , {(a^\dagger_{s'})^k}_l \Big\} = \delta^s_{s'} \delta_{i}^k \delta^{j}_l
\end{equation}
Moreover, all raising operators in $B_s$ commute with all raising operators in $B_s$ and $F_s$ while all raising operators in $F_s$ anticommute with each other.
With this, one can define the Hilbert spaces $\CH'$ and $\CH$ from Eqs.~\eqref{thebasisprime}, \eqref{singletcond} and \eqref{thebasis}. Specifically, the Hilbert space $\CH$ of Spin Matrix theory is the linear space spanned by the states \eqref{thebasis}, or equivalently \eqref{tracebasis}.

The split up of the spin group representation $R_s = B_s \oplus F_s$ into bosonic and fermionic excitations happens for instance for representations of Lie supergroups of the type $SU(p,q|r)$ with both $p+q$ and $r$ non-zero. Here the generators in the $su(p,q)$ and $su(r)$ subalgebras of the $su(p,q|r)$ algebra are bosonic while the remaining generators are fermionic. While a bosonic generator acting on $s \in B_s$ gives an element in $B_s$ a fermionic generator acting on $s$ gives an element in $F_s$ and so forth.

\subsubsection*{Hamiltonian of Spin Matrix theory}

We consider now interactions in Spin Matrix theory. The type of interaction that we consider is where two excitations are annihilated and two new are created. We demand furthermore that the interaction should commute with all generators of the spin group $G_s$ and that the spin and matrix parts factorize. The general form of such a Hamiltonian is
\begin{equation}
\label{Hint1}
H_{\rm int} = \frac{1}{N} U_{s r}^{s' r'} \sum_{\sigma\in S(4)} T_{\sigma} \, {(a^\dagger_{s'})^{i_{\sigma(1)}}}_{i_{3}} {(a^\dagger_{r'})^{i_{\sigma(2)}}}_{i_{4}}  {(a^{s})^{i_{\sigma(3)}}}_{i_{1}} {(a^{r})^{i_{\sigma(4)}}}_{i_{2}}
\end{equation}
where $T_{\sigma}$, $\sigma \in S(4)$, are coefficients and where the sums over $s,r,s',r'$ and $i_1,i_2,i_3,i_4$ are
understood (note that the factor of $1/N$ is for later convenience). 
One can check that this Hamiltonian preserves the singlet condition \eqref{singletcond} and hence stays within the Hilbert space $\CH$. Furthermore, it is a Hermitian operator on $\CH$ provided the spin part is a Hermitian matrix $(U^{s' r'}_{sr})^* = U^{sr}_{s'r'}$ and that the $T_{\sigma}$ coefficients obey $T_{\sigma^{-1}} = T_{\sigma}$.
We choose the $T_\sigma$ coefficients for Spin Matrix theory such that
\begin{equation}
\label{sigmaT}
\sum_{\sigma \in S(4)} T_\sigma \sigma =(14) + (23) - (12) - (34) 
\end{equation}
We take this explicit choice of $T$ since it describes the behavior near zero-temperature critical points of $\CN=4$ SYM, as we shall see below. Furthermore, it ensures that the Hamiltonian reduces to that of a general nearest-neighbor spin chain for $N=\infty$, as we shall see in Section \ref{sec:SCfromSMT}.

Turning now to the spin part of the interaction in \eqref{Hint1} we see that
$U$ is a linear operator which takes an element in $R_s \otimes R_s$ and gives a new element in $R_s \otimes R_s$
\begin{equation}
U: R_s \otimes R_s \rightarrow R_s \otimes R_s
\end{equation}
We see
from the form of $H_{\rm int}$ in \eqref{Hint1} that $U_{s r}^{s' r'} =
U_{r s}^{r' s'}$. Expand now the product representation $R_s \otimes R_s$ into irreducible representations
\begin{equation}
\label{productRs}
R_s \otimes R_s = \sum_{\CJ}  V_{\CJ}
\end{equation}
where $\CJ$ labels the irreducible representations $V_\CJ$ (labelling includes multiplicities).
We impose that $H_{\rm int}$ should commute with all generators of the spin group $G_s$.
This means that in each subspace $V_{\CJ}$ the interaction $U$ is
proportional to the identity matrix, hence
\begin{equation}
\label{Uform}
U_{s r}^{s' r'} = \sum_{\CJ} C_{\CJ} (P_{\CJ})_{s r}^{s' r'}
\end{equation}
where $P_{\CJ}$ is the projector that projects from $R_s \otimes
R_s$ into $V_{\CJ}$ for a given ${\CJ}$. We see thus that the only
freedom in choosing the interaction $H_{\rm int}$ lies in choosing
the constants $C_{\CJ}$.

In general we include also a diagonal piece in the Hamiltonian. Define the operator
\begin{equation}
\label{Lop}
L = \sum_s \tr ( a_s^\dagger a^s )
\end{equation}
This gives what we call the {\sl length} of a state and it commutes with the generators of $G_s$ ($e.g.$ for the state in \eqref{tracebasis} the length is $L$). In addition we have the Cartan generators of $G_s$ here denoted $K_p$ with $p$ labelling them. Thus, we take our most general Hamiltonian to be of the form
$H = g H_{\rm int} + \mu_0 L - \sum_p \mu_p K_p$.
One notices that the partition function at a temperature $T$ is invariant under the rescaling $T \rightarrow \alpha T$, $g \rightarrow \alpha g$, $\mu_0 \rightarrow \alpha \mu_0$ and $\mu_p \rightarrow \alpha \mu_p$ and hence one can remove a parameter. One could choose $g=1$ which would connect high (low) temperature to weak (strong) coupling. However, we choose instead $\mu_0=1$ since then we can connect high (low) temperature to long (short) average lengths of the states. Since non-planar effects increase with the length one gets that for low temperature the theory becomes effectively planar  (assuming $N$ is large) and for very high temperature the theory is highly non-planar. 
In summary, the Hamiltonian of Spin Matrix theory is
\begin{equation}
\label{theH}
H = L + g H_{\rm int} - \sum_p \mu_p K_p
\end{equation}
where $g$ is the coupling constant of the interaction and $\mu_p$ can be regarded as chemical potentials. 
Hence we can write the partition function for Spin Matrix theory as
\begin{equation}
\label{theZ}
Z (\beta, \mu_p) = \tr ( e^{-\beta H} ) = \tr ( e^{-\beta (L +  g H_{\rm int}  - \sum_p  \mu_p K_p) } ) 
\end{equation}
where the trace is over the Hilbert space $\CH$.

%%%%%%%%%%%%%%%%%%%%%%%%%%%%%%%%%%%%%%%%%%%%%%%%
\subsection{Spin chains from Spin Matrix theory}
\label{sec:SCfromSMT}

We consider here Spin Matrix theory in the planar limit $N=\infty$. In the planar limit $N=\infty$ the multi-trace states \eqref{tracebasis} are linearly independent and provide therefore a basis. The Hilbert space $\CH$ of Spin Matrix theory can thus be thought of as being that of a gas of single trace states. Consider a single-trace state 
\begin{equation}
\label{singletrace}
| s_1 s_2 \cdots s_L \rangle \equiv \tr ( a^\dagger_{s_1} a^\dagger_{s_2}  \cdots a^\dagger_{s_L} ) | 0 \rangle
\end{equation}
One can interpret this as a spin chain with translation invariance (due to the cyclicity of the trace) since the contraction between the individual raising operator clearly defines a succession of the spins \cite{Minahan:2002ve}. Note that having $N=\infty$ is crucial for the spin chain interpretation. If one has $L > N$ one can generically write the single trace as a linear combination of multi-trace states of the type \eqref{tracebasis}, all built from single-traces shorter than $L$, and hence the succession of the spins is no longer well-defined.%
\footnote{On the other hand, we show in Section \ref{sec:suq_partfcts} that for a generic Spin Matrix theory Hilbert space a given basis of multi-trace states necessarily contains arbitrarily long single-traces.} 

Consider next the action of the interacting part of the Hamiltonian $H_{\rm int}$ on a contracted two-oscillator state 
\begin{equation}
\label{Hinttwoosc}
H_{\rm int} {(a^\dagger_{m})^{i}}_{j} {(a^\dagger_{n})^{j}}_{l}  | 0 \rangle = \frac{2}{N}
U^{rs}_{mn} [   \delta^i_l {(a^\dagger_{r})^{i'}}_{j} {(a^\dagger_{s})^{j}}_{i'}  + N {(a^\dagger_{r})^{i}}_{j'} {(a^\dagger_{s})^{j'}}_{l} - {(a^\dagger_{r})^{j}}_{j} {(a^\dagger_{s})^{i}}_{l} - {(a^\dagger_{r})^{i}}_{l} {(a^\dagger_{s})^{j}}_{j}   ] |0\rangle
\end{equation}
One can see that the second term dominates for large $N$ since it is proportional to $N$. Thus, $H_{\rm int}$ has an extra factor of $N$ when one applies it to two contracted oscillators. The planar limit $N\rightarrow \infty$ of $H_{\rm int}$ can only be non-singular if $U^{rs}_{mn}$ is finite in the limit. In fact, we assume that $U^{rs}_{mn}$ does not depend on $N$, $i.e.$ that the coefficients $C_\CJ$ are independent of $N$. This means that in the planar limit only the action of $H_{\rm int}$ on the contracted oscillators survive. Hence $H_{\rm int}$ is non-zero only when applied to nearest-neighbor sites on a single-trace, and zero (going to zero as $1/N$) when applied to two oscillators belonging to two different single-traces (say in a multi-trace state) or when applied to two non-nearest neighbor operators in a single trace.
The action of $H_{\rm int}$ on \eqref{singletrace} is
\begin{equation}
\label{Hintspin}
H_{\rm int} | s_1 s_2 \cdots s_L \rangle
=  2  \sum_{k=1}^L U^{mn}_{s_k s_{k+1}}  | s_{1} \cdots s_{k-1}\, m\, n\, s_{k+2}  \cdots s_L  \rangle
\end{equation} 
Thus on a single trace state the Hamiltonian is given essentially by $U$ in \eqref{Uform} acting on neighboring spins. This is thus a nearest neighbor spin chain Hamiltonian. 
We see from this that the Spin Matrix theory has a unique extension from the $N=\infty$ limit to the full finite $N$ theory since the full Spin Matrix theory is uniquely determined by $U^{rs}_{pq}$.

As already remarked, the multi-trace state basis \eqref{tracebasis} can be interpreted as the basis for a gas of spin chains with spin chain Hamiltonian \eqref{Hintspin}. We can thus say that in the planar limit the partition function of Spin Matrix theory is that of a gas of spin chains. 
Relaxing the planar limit to large $N$ one can still effectively regard the states \eqref{tracebasis} as a basis for low enough temperatures and energies such that the average length of a multi-trace is smaller than $N$. However, the other terms in $H_{\rm int}$ that go like $1/N$ will now be non-zero and enable that the spin chains can split or join with a propability of order $1/N$. Thus, assuming large $N$, one concludes that at low energy/temperatures Spin Matrix theory can be thought of as a gas of weakly interacting spin chains. We shall exhibit this in greater detail for a specific Spin Matrix theory below.

%%%%%%%%%%%%%%%%%%%%%%%%%%%%%%%%%%%%%%%%%%%%%%%%%%%%%%%%%%
\section{Spin Matrix theory from $\CN=4$ SYM near critical points}
\label{sec:SMTlimit}

In this section we show that Spin Matrix theory describes $\CN=4$ SYM near zero-temperature critical points in the grand canonical ensemble. We begin by reviewing the partition function of $\CN=4$ SYM. We define then our notion of zero-temperature critical points in the grand canonical ensemble of $\CN=4$ SYM. There are nine critical points and we show how Spin Matrix theory emerges near these. Finally we show that Spin Matrix theory equivalently can be seen to emerge in the microcanonical ensemble in a low energy and non-relativistic limit.

%%%%%%%%%%%%%%%%%%%%%%%%%%%%%%%%%%%%%%%%%%%%%%%%%%%%%%%%%%
\subsection{Partition function of $\CN=4$ SYM}
\label{sec:SYM}

Consider $\CN=4$ SYM on $\R\times S^3$ with gauge group $U(N)$. This theory has global symmetry $PSU(2,2|4)$. The bosonic subgroup $SU(2,2)\simeq SO(2,4)$ has Cartan generators being the dilatation operator $D$, and the two angular momenta on $S^3$ called $S_1$ and $S_2$. The bosonic subgroup $SU(4)\simeq SO(6)$ has the three R-symmetry generators $R_1$, $R_2$ and $R_3$ here chosen as Cartan generators of $SO(6)$. The grand canonical partition function is
\begin{equation}
\label{sympartfct}
Z(\beta,\vec{\Omega}) = \tr \left( e^{-\beta D + \beta \vec{\Omega} \cdot \vec{J}} \right)
\end{equation} 
at temperature $T=1/\beta$, chemical potentials $\vec{\Omega} = ( \omega_1,\omega_2, \Omega_1,\Omega_2,\Omega_3)$ and 't Hooft coupling $\lambda=\gym^2 N$. In addition to the notation $\vec{\Omega}$ for the chemical potentials we also introduce $\vec{J}= (S_1,S_2,R_1,R_2,R_3)$ and $\vec{\Omega} \cdot \vec{J} = \omega_1 S_1 + \omega_2 S_2 + \Omega_1 R_1 + \Omega_2 R_2 + \Omega_3 R_3$. The trace in \eqref{sympartfct} is over the operators of $\CN=4$ SYM on $\R^4$  (or corresponding states of $\CN=4$ SYM on $\R \times S^3$). These are all the operators spanned by the multi-trace operators built out of the letters of $\CN=4$ SYM transforming in the adjoint representation of $U(N)$. Seen from the point of view of states of $\CN=4$ SYM on $\R \times S^3$ the reason for the singlet condition ($i.e.$ that there is no free $U(N)$ indices) is that one cannot have a net charge on a three-sphere since flux lines of a charge need to end somewhere \cite{Sundborg:1999ue,Aharony:2003sx}.

In general the dilatation operator can be written as $D=D_0 + \delta D$ where $D_0 = D|_{\lambda=0}$ and $\delta D$ is the anomalous dimension part. At one-loop we write $\delta D = \lambda D_2 + \CO(\lambda^{3/2})$. The $D_2$ operator acts on two letters at a time, each letter being in the singleton representation $\CA$ of $psu(2,2|4)$. The product of two singleton representations is $\CA \otimes \CA = \sum_{j=0}^\infty \CV_j$ where $\CV_j$ are irreducible representations labelled uniquely by the quadratic Casimir of $psu(2,2|4)$ (see \cite{Dolan:2002zh,Bianchi:2003wx} for details). Using this the one-loop dilatation operators has the form \cite{Beisert:2003jj}
\begin{equation}
\label{d2op} D_2 = - \frac{1}{8\pi^2 N} \sum_{j=0}^\infty h(j)
(P_j)^{AB}_{CD} : \mbox{Tr} [ W_A , \partial_{W_C} ] [W_B , \partial_{W_D}] :
\end{equation}
where $h(j) = \sum_{k=1}^j \frac{1}{k}$ are the harmonic numbers ($h(0)=0$), $P_j$ is the projection operator from $\CA \otimes \CA$ to $\CV_j$, $W_A$ with $A\in \CA$ represents all possible letters of $\CN=4$ SYM and one has normal ordering such that $\partial_{W}$ is moved to the right of $W$.

If one could artificially remove the interactions of $\CN=4$ SYM beyond one-loop one could recast the resulting theory as a Spin Matrix theory corresponding to the representation $\CA$ of the group $PSU(2,2|4)$. For the Hilbert space one simply identifies the matrix valued raising operator $a^\dagger_{s}$ with the letter $W_s$ for all $s\in \CA$. This gives a one-to-one map between the gauge theory operators in $\CN=4$ SYM and the Hilbert space of the Spin Matrix theory. Turning to the one-loop dilatation operator \eqref{d2op} this is equal to $H_{\rm int}$ if we identify the label $\CJ$ with $j$, the representations $V_\CJ$ with $\CV_j$ and
\begin{equation}
\label{Cj}
C_j = \frac{1}{8\pi^2} h(j)  \ , \ j=0,1,2,...
\end{equation}
However, in the end $\CN=4$ SYM cannot be a Spin Matrix theory since $\CN=4$ SYM is a quantum field theory and hence should have local relativistic invariance, including particle-antiparticle creation, and hence it cannot be identified with a non-relativistic quantum mechanical theory. Indeed, the above naive truncation of the higher-loop terms - keeping only the one-loop term - is unphysical and the relativistic behavior of $\CN=4$ SYM is precisely a consequence of having the full dilatation operator, as one can see for example in the planar limit from the dispersion relation for a single magnon \cite{Beisert:2005tm} as well as in the BMN limit \cite{Berenstein:2002jq}. Instead, as we shall see below, the limits in which one obtains Spin Matrix theory from $\CN=4$ SYM involve $\lambda \rightarrow 0$ and gives a natural way to only keep the one-loop term of the dilatation operator. This is tied to the fact that the Spin Matrix theory limit is non-relativistic, and in addition one is naturally restricted to a subsector of the space of operators, which further simplifies the theory in comparison to $\CN=4$ SYM.

%%%%%%%%%%%%%%%%%%%%%%%%%%%%%%%%%%%%%%%%%%%%%%%%%%%%%%%%%%
\subsection{Zero-temperature critical points in the grand canonical ensemble}
\label{sec:critpts}

Consider $\CN=4$ SYM on $\R \times S^3$ at large $N$ in the grand canonical ensemble parametrized by $(T,\vec{\Omega})$. For any coupling $\lambda$ and zero chemical potentials one has a definite temperature at which there is a phase transition from confining behavior $\log Z \sim \CO (1)$ to deconfining behavior $\log Z \sim \CO(N^2)$ of the partition function. This phase transition persists also for non-zero chemical potentials $\vec{\Omega}$ thus defining a submanifold of transition points in the grand canonical ensemble $(T,\vec{\Omega})$. 
We define the {\sl zero-temperature critical points} of the grand canonical ensemble as the points that one can obtain by continuing this submanifold of phase transition points to zero temperature. Thus, for a given critical point $(T,\vec{\Omega})=(0,\vec{\Omega}^{(c)})$ there are confinement/deconfinement transition points that lie arbitrarily close to it. In Table \ref{tab:SMT} we listed nine critical points for $\CN=4$ SYM.

\begin{table}[ht]
\begin{center}
\begin{tabular}{|c||c|c|c|}
\hline Critical point   & Spin group  & Cartan diagram  & Representation   \\
$(T,\omega_1,\omega_2,\Omega_1,\Omega_2,\Omega_3)$ & $G_s$ &  for algebra & $R_s$ \\ 
\hline 
$(0,0,0,1,1,0)$ & $SU(2)$ & $\bigcirc$ & $[1]$\\
\hline
$(0,\frac{2}{3},0,1,\frac{2}{3},\frac{2}{3})$ & $SU(1|1)$ & $\bigotimes$ &$[1]$ \\
\hline
$(0,\frac{1}{2},0,1,1,\frac{1}{2})$ & $SU(1|2)$ & $\bigcirc \!\!-\!\!\!-\!\! \textstyle \bigotimes$ &$[1,0]$ \\
\hline
$(0,0,0,1,1,1)$ & $SU(2|3)$ & $\bigcirc \!\!-\!\!\!-\!\! \textstyle \bigotimes  \!\!-\!\!\!-\!\! \bigcirc \!\!-\!\!\!-\! \bigcirc$ &$[0,0,0,1]$ \\
\hline
$(0,1,0,1,0,0)$ & $SU(1,1)$ &  $\bigcirc$ &$[-1]$ \\
\hline
$(0,1,0,1,\frac{1}{2},\frac{1}{2})$ & $SU(1,1|1)$ & $\textstyle \bigotimes \!\!-\!\!\!-\!\! \textstyle \bigotimes$ &$[0,1]$ \\
\hline
$(0,1,0,1,1,0)$ & $SU(1,1|2)$ & $\textstyle \bigotimes \!\!-\!\!\!-\!\! \bigcirc \!\!-\!\!\!-\!\! \textstyle \bigotimes$ & $[0,1,0]$\\
\hline
$(0,1,1,1,0,0)$ & $SU(1,2|2)$ & $\bigcirc \!\!-\!\!\!-\!\!  \textstyle \bigotimes \!\!-\!\!\!-\!\!
\bigcirc \!\!-\!\!\!-\!\! \textstyle \bigotimes
$ & $[0,0,0,1]$\\
\hline
$(0,1,1,1,1,1)$ & $SU(1,2|3)$ & $\bigcirc \!\!-\!\!\!-\!\!  \textstyle \bigotimes \!\!-\!\!\!-\!\!
\bigcirc \!\!-\!\!\!-\!\! \bigcirc \!\!-\!\!\!-\!\! \textstyle \bigotimes$
 &$[0,0,0,1,0]$ \\
\hline
\end{tabular}
\caption{Critical points of $\CN=4$ SYM that can be described by Spin Matrix theory. Listed are the spin groups, the Cartan diagram for the corresponding algebra and the representations (in terms of Dynkin labels) that defines the Spin Matrix Theory for a given critical point. 
 \label{tab:SMT}} 
\end{center}
 \end{table}

One can see from the partition function \eqref{sympartfct} that a necessary requirement for a critical point $(0,\vec{\Omega}^{(c)})$ is that $D \geq \vec{\Omega}^{(c)} \cdot \vec{J}$ for all operators of $\CN=4$ SYM while at the same time there should exist operators that saturate the bound. We restrict ourselves to critical points for which there are protected operators that saturate the bound, indeed all the critical points of Table \ref{tab:SMT} are of this type.%
\footnote{To make the list of this type of critical points complete one should include the fact that the $SU(1|1)$ point is part of a larger family of critical points $(0,a,-b,1,1-\frac{1}{2}(a+b),1-\frac{1}{2}(a+b))$, $0 < a,b < 1$ and that one has another $SU(1|2)$ point $(0,\frac{1}{2},\frac{1}{2},1,1,0)$ as well as another $SU(1,1|1)$ point $(0,1,\frac{1}{2},1,\frac{1}{2},0)$. In addition there are equivalent critical points obtained by interchanging $\omega_1$ and $\omega_2$, or by permuting $\Omega_1$, $\Omega_2$ and $\Omega_3$.}
 Then we can infer from the results of \cite{Harmark:2007px} that the above requirement is sufficient as well.%
\footnote{With this restriction the critical points considered here are also critical points in the sense of \cite{Harmark:2007px}, $i.e.$ in \cite{Harmark:2007px} we defined critical points $(0,\vec{\Omega}^{(c)})$ to be such that $D_0 \geq \vec{\Omega}^{(c)} \cdot \vec{J}$ for all operators of $\CN=4$ SYM while at the same time there should exist operators that saturate the bound.}

We now examine $\CN=4$ SYM with partition function \eqref{sympartfct} as one approaches one of the critical points of Table \ref{tab:SMT}. Writing the critical point as $(0,\vec{\Omega}^{(c)})$ we are taking the limit $(T,\vec{\Omega}) \rightarrow (0,\vec{\Omega}^{(c)})$. We require that $\beta ( \vec{\Omega}-\vec{\Omega}^{(c)})$ is finite in the limit.
We record the identity
\begin{equation}
\label{theidentity}
\beta D - \beta \vec{\Omega} \cdot \vec{J} =  \beta \delta D + \beta (D_0  -  \vec{\Omega}^{(c)} \cdot \vec{J}) - \beta ( \vec{\Omega}-\vec{\Omega}^{(c)} ) \cdot \vec{J}
\end{equation}
We first analyze the limit for $\lambda=0$ hence $\delta D=0$.
Consider $\Delta \equiv D_0  -  \vec{\Omega}^{(c)} \cdot \vec{J}$ for the states of $\CN=4$ SYM on $\R \times S^3$. Either $\Delta=0$ or $\Delta  \geq 1/2$ hence only states with $\Delta =0 $ contributes to the partition function after the limit. One finds that the $\Delta =0$ states correspond to the Hilbert space $\CH$ of Spin Matrix theory \eqref{tracebasis} with the spin group $G_s$ being a subgroup of $PSU(2,2|4)$ and the representation $R_s$ being a subset of $\CA$ \cite{Harmark:2007px}. In Table \ref{tab:SMT} we listed the representations $R_s$ and the groups $G_s$ corresponding to each of the nine critical points.

Considering further the limit towards the critical point for $\lambda=0$ one can show that the term $- \beta ( \vec{\Omega}-\vec{\Omega}^{(c)} ) \cdot \vec{J}$ in \eqref{theidentity} in general is a linear combination of the length operator of Eq.~\eqref{Lop} and the Cartan generators of $G_s$ denoted $K_p$ for the $\Delta=0$ states \cite{Harmark:2007px}. Hence 
\begin{equation}
- \beta ( \vec{\Omega}-\vec{\Omega}^{(c)} ) \cdot \vec{J} \rightarrow  \tilde{\beta} \Big(  L - \sum_p \mu_p K_p \Big)
 \end{equation}
for the $\Delta=0$ states in the limit $(T,\vec{\Omega}) \rightarrow (0,\vec{\Omega}^{(c)})$ with appropriate choices of $\tilde{\beta}$ and $\mu_p$. 
 
Turning on $\lambda$ we get the additional term $\beta \delta D$ in \eqref{theidentity}. If we keep $\lambda$ fixed and non-zero in the $\beta \rightarrow \infty$ limit we clearly get the further condition $\delta D =0$ on the states. Hence we only get contributions to the partition function from supersymmetric states.
To get an interacting theory we should send $\lambda \rightarrow 0$ with $\beta\rightarrow \infty$ such that $\beta \lambda$ is finite in the limit. One can write this as $\beta \lambda\rightarrow \tilde{\beta} g$ where we introduced the finite parameter $g$. 
Then $\beta \delta D \rightarrow \tilde{\beta}\tilde{\lambda} D_2$ since the higher loop terms in $\delta D$ go to zero. For the $\Delta=0$ states, which can be seen as states in the Hilbert space $\CH$ of the Spin Matrix theory corresponding to the representation $R_s$ of the group $G_s$ recorded in Table \ref{tab:SMT}, we have $D_2 = H_{\rm int}$ with the identification \eqref{Cj}. However, since $R_s$ is not $\CA$ but instead the representations given in Table \ref{tab:SMT} one should be careful in interpreting the label $j$ in \eqref{d2op} and \eqref{Cj}. To this end, we record that for a highest weight state one has \cite{Beisert:2004ry}
\begin{equation}
\label{casimirsu224}
j(j+1)= \frac{1}{2}D_0^2 + 2D_0 + \frac{1}{2} ( S_1^2+ S_2^2 ) - S_1 - \frac{1}{2} (R_1^2 + R_2^2 + R_3^2) - 2R_1 - R_2
\end{equation}
For instance, for the $SU(2)$ critical point of Table \ref{tab:SMT} $R_s$ is the spin $1/2$ representation and we can label the irreducible representations $V_\CJ$ in \eqref{productRs} using the casimir $s(s+1)$ of the $SU(2)$ algebra (being either  spin 0 ($s=0$) or spin 1 ($s=1$) representation). Hence $\CJ=s$ in this case. One can check using Eq.~\eqref{casimirsu224} that the spin 0 representation corresponds to $j=1$ while the spin 1 representation to $j=0$. Hence, using the label $\CJ=s$ in \eqref{productRs} we get for the coefficients $C_s$, $s=0,1$, 
\begin{equation}
\label{su2coeff}
C_{s=0} = \frac{1}{8\pi^2} \spa C_{s=1} = 0
\end{equation}
from \eqref{Cj}. We examine this particular Spin Matrix theory, which we dub $SU(2)$ Spin Matrix theory,  below in Sections \ref{sec:lowT} and \ref{sec:hightemp}.

We conclude that approaching one of the critical points $(0,\vec{\Omega}^{(c)})$ listed in Table \ref{tab:SMT} in the limit
\begin{equation}
\label{SMTlim}
(T,\vec{\Omega}) \rightarrow (0,\vec{\Omega}^{(c)}) \ \ \mbox{and} \ \ \lambda \rightarrow 0 \ \ \ \ \mbox{with}\ \ \ \ \beta ( \vec{\Omega}-\vec{\Omega}^{(c)}) \ \ \mbox{and}\ \  \beta \lambda \ \ \mbox{finite}
\end{equation}
of $\CN=4$ SYM one finds the partition function 
\begin{equation}
\label{theZ2}
Z (\tilde{\beta}, \mu_p)  = \tr ( e^{-\tilde{\beta} ( L +  g H_{\rm int}  -   \sum_p \mu_p K_p )} ) 
\end{equation}
which is the partition function of Spin Matrix theory with the spin group $G_s$, along with the representation $R_s$ thereof, as recorded in Table \ref{tab:SMT}. Moreover, the coefficients in the interaction term $H_{\rm int}$ are given by Eq.~\eqref{Cj}. The trace in the partition function is over the Hilbert space $\CH$ of Eq.~\eqref{tracebasis} corresponding to the subsector $\Delta=0$ of $\CN=4$ SYM.

%%%%%%%%%%%%%%%%%%%%%%%%%%%%%%%%%%%%%%%%%%%%%%%%%%%%%%%%%%
\subsection{Low energy and non-relativistic limits (microcanonical ensemble)}
\label{sec:microlim}

One can find equivalent limits of $\CN=4$ SYM on $\R\times S^3$ in the microcanonical ensemble. Given a critical point $(0,\vec{\Omega}^{(c)})$ of Table \ref{tab:SMT} one has $D \geq \vec{\Omega}^{(c)} \cdot \vec{J}$ for all states. Therefore, it makes sense to take a low energy limit $D - \vec{\Omega}^{(c)} \cdot \vec{J} \rightarrow 0$. This means that the states above the energy gap $\Delta \geq 1/2$ (defining again $\Delta \equiv D_0  -  \vec{\Omega}^{(c)} \cdot \vec{J}$) effectively decouple and one is left with the $\Delta=0$ states which correspond to the states in the Hilbert space of Spin Matrix theory with spin group $G_s$ and spin label in the representation $R_s$ as given in Table \ref{tab:SMT}. For states with $\Delta =0$ we have $D - \vec{\Omega}^{(c)} \cdot \vec{J} = \delta D = \lambda D_2 + \CO( \lambda^{3/2} )$. Hence to get a non-trivial energy spectrum we should take the limit
\begin{equation}
\label{SMTmicrolim}
D - \vec{\Omega}^{(c)} \cdot \vec{J} \rightarrow 0 \ \ \mbox{and} \ \ \lambda \rightarrow 0 \ \ \ \ \mbox{with}\ \ \ \ \frac{D - \vec{\Omega}^{(c)} \cdot \vec{J}}{\lambda} \ \ \mbox{finite}
\end{equation}
which gives Spin Matrix theory with interaction $H_{\rm int} = (D - \vec{\Omega}^{(c)} \cdot \vec{J})/\lambda$ 
with the spin group $G_s$, along with the representation $R_s$ thereof, as recorded in Table \ref{tab:SMT}, and with the coefficients in the interaction term $H_{\rm int}$ given by Eq.~\eqref{Cj}.
The limit \eqref{SMTmicrolim} is equivalent to \eqref{SMTlim}. In the microcanonical ensemble one has, in addition to $H_{\rm int}$, the length operator $L$ as well as the Cartan operators $K_p$ held fixed. Hence one can go to the grand canonical ensemble of Spin Matrix theory with partition function \eqref{theZ2} after the limit \eqref{SMTmicrolim}.

While we established the limits \eqref{SMTlim} and \eqref{SMTmicrolim} as low energy limits in which part of the spectrum of the states of $\CN=4$ SYM on $\R\times S^3$ decouple we point out that they in addition can be seen as non-relativistic limits. This one can see in the planar limit by considering a magnon of the $psu(2,2|4)$ spin chain for $\CN=4$ SYM which has dispersion relation $\delta D = \sqrt{1+\frac{\lambda}{\pi^2} \sin^2 \frac{p}{2} } - 1$ where $p$ is the momentum of the magnon on the spin chain \cite{Beisert:2005tm}. For small momenta this becomes the relativistic dispersion relation of a free particle. Instead when taking the limit $\lambda \rightarrow 0$ we get a non-relativistic dispersion relation for small momenta. For the $SU(2)$ critical point of Table \ref{tab:SMT} this limit from relativistic to non-relativistic symmetry is considered in detail in \cite{Harmark:2008gm}.

%%%%%%%%%%%%%%%%%%%%%%%%%%%%%%%%%%%%%%%%%%%%%%%%%%%%%%%%%%
\subsection{The $g \rightarrow \infty$ limit, supersymmetry and the AdS/CFT correspondence}
\label{sec:g_infty}

In this section we consider two limits that both end at one of the zero-temperature critical points of Table \ref{tab:SMT}. Employing a conjecture regarding supersymmetric states of $\CN=4$ SYM we can use these limits to show that the $g\rightarrow \infty$ limit of Spin Matrix theory matches the string theory side of the AdS/CFT correspondence.

In Section \ref{sec:critpts} we find that approaching one of the zero-temperature critical point of Table \ref{tab:SMT} $(T,\vec{\Omega}) \rightarrow (0,\vec{\Omega}^{(c)})$ with $\lambda$ fixed and non-zero, one obtains the condition $\delta D=0$ which only holds for supersymmetric states. More precisely, the $\CN=4$ SYM partition function in the limit
\begin{equation}
(T,\vec{\Omega}) \rightarrow (0,\vec{\Omega}^{(c)}) \ \ \ \ \mbox{with}\ \ \ \ \beta ( \vec{\Omega}-\vec{\Omega}^{(c)}) \ \ \mbox{finite} \ \ \mbox{and}\ \   \lambda > 0 \ \ \mbox{fixed}
\end{equation}
gives the partition function
\begin{equation}
\label{partlim1}
Z(\tilde{\beta},\mu_p) =  \tr ( e^{-\tilde{\beta} ( L  -   \sum_p \mu_p K_p )} ) \ \ \ \ \ \ \mbox{(Trace over SUSY states)}
\end{equation}
Alternatively, one can take first the Spin Matrix theory limit \eqref{SMTlim}, and then subsequently take the limit $g\rightarrow \infty$ of the Spin Matrix theory, $i.e.$ 
\begin{equation}
\begin{array}{l}\ds
\mbox{Step 1:}\ \ \ (T,\vec{\Omega}) \rightarrow (0,\vec{\Omega}^{(c)}) \ \ \mbox{and} \ \ \lambda \rightarrow 0 \ \ \ \ \mbox{with}\ \ \ \ \beta ( \vec{\Omega}-\vec{\Omega}^{(c)}) \ \ \mbox{and}\ \  \beta \lambda \ \ \mbox{finite} \\[2mm]\ds
\mbox{Step 2:}\ \ \  g \rightarrow \infty
\end{array}
\end{equation}
Then we obtain the partition function 
\begin{equation}
\label{partlim2}
Z(\tilde{\beta},\mu_p) =  \tr ( e^{-\tilde{\beta} ( L  -   \sum_p \mu_p K_p )} ) \ \ \ \ \ \ \mbox{(Trace over $H_{\rm int}=0$ states)}
\end{equation}
The two partition functions \eqref{partlim1} and \eqref{partlim2} are in fact the same partition function, provided that it is true that $D_2=0$ is equivalent to $\delta D=0$ (for a non-zero $\lambda$) for states with $D_0 = \vec{\Omega}^{(c)} \cdot \vec{J}$. That $\delta D=0$ implies $D_2=0$ is easy to see since $\delta D$ in fact implies all loop orders of the dilatation operator are zero. Instead the reverse statement is non-trivial. However, this is conjectured to be true \cite{Kinney:2005ej,Berkooz:2006wc,Grant:2008sk} and has been confirmed for $1/8$ BPS states \cite{Kinney:2005ej,Biswas:2006tj}. We assume here the validity of this conjecture, thus also for $1/16$ BPS states.

A consequence of the two partition functions \eqref{partlim1} and \eqref{partlim2} being equal is that we can use the $g\rightarrow \infty$ limit of Spin Matrix theory to compute the partition function \eqref{partlim1} for any non-zero $\lambda$. Taking in particular $\lambda \gg 1$ this partition function can be mapped by the AdS/CFT correspondence to the corresponding partition function on the string theory side. This  is obtained in the following limit of string theory in the grand canonical ensemble (dual to the grand canonical ensemble of $\CN=4$ SYM) 
\begin{equation}
\label{stringlim}
(T,\vec{\Omega}) \rightarrow (0,\vec{\Omega}^{(c)}) \ \ \ \ \mbox{with}\ \ \ \ \beta ( \vec{\Omega}-\vec{\Omega}^{(c)}) \ \ \mbox{finite} \ \ \mbox{and}\ \   g_s, \ N \ \ \mbox{fixed}
\end{equation}
and gives 
\begin{equation}
\label{partlim3}
Z(\tilde{\beta},\mu_p) =  \tr ( e^{-\tilde{\beta} ( L  -   \sum_p \mu_p K_p )} ) \ \ \ \ \ \ \mbox{(String theory)}
\end{equation}
with the trace being over the supersymmetric states on the string theory side that survive the limit \eqref{stringlim}.%
\footnote{Note that one can formulate the above just as well in the microcanonical ensemble as we did for $\CN=4$ SYM in Section \ref{sec:microlim}, hence as a low energy and non-relativitistic limit of string theory.}

In conclusion, we have shown that we can quantitatively match the $g\rightarrow \infty$ limit of Spin Matrix theory, corresponding to one of the zero-temperature critical points of Table \ref{tab:SMT}, to the limit \eqref{stringlim} of string theory on $\ads_5\times S^5$. This means in particular that for the supersymmetric sector we can map finite-$N$ effects of Spin Matrix theory to non-perturbative effects in string theory, in line with our philosophy illustrated in Figure \ref{fig:philosophy}.

%%%%%%%%%%%%%%%%%%%%%%%%%%%%%%%%%%%%%%%%%%%%%%%%%%%%%%%%%%
\section{$SU(2)$ Spin Matrix theory at low temperature (nearly planar)}
\label{sec:lowT}

In this section, as well as in Section \ref{sec:hightemp}, we explore the phases of $SU(2)$ Spin Matrix theory, as illustrated in Figure \ref{fig:SMT}. We begin below by writing down the Hamiltonian of $SU(2)$ Spin Matrix theory in detail in Section \ref{sec:defSMTsu2}. We subsequently review the low temperature phases in Sections \ref{sec:planarlimit} and \ref{sec:nearlyplanar}.

\subsection{$SU(2)$ Spin Matrix theory}
\label{sec:defSMTsu2}

$SU(2)$ Spin Matrix theory is the theory that one obtains near the critical point $(T,\vec{\Omega})=(0,0,0,1,1,0)$ as listed in Table \ref{tab:SMT}. This has spin group $G_s= SU(2)$. The representation $R_s = \frac{1}{2}$ is the fundamental spin $1/2$ representation. We label this as spin-up and spin-down $s=\uparrow,\downarrow$. This specifies the Hilbert space $\CH$. The interacting Hamiltonian $H_{\rm int}$ is given by the coefficients \eqref{su2coeff}. From this we read that the spin part of the Hamiltonian $U^{s'r'}_{sr}$ is proportional to the projector from $\frac{1}{2} \otimes \frac{1}{2}$ to the spin 0 representation. This projector is $(P_{s=0})^{s'r'}_{sr}=\frac{1}{2}( \delta^{r'}_r \delta^{s'}_s - \delta^{r'}_s \delta^{s'}_r )$ and hence
\begin{equation}
\label{U_su2}
U^{s'r'}_{sr} = \frac{1}{16\pi^2}  ( \delta^{r'}_r \delta^{s'}_s - \delta^{r'}_s \delta^{s'}_r )
\end{equation}
Inserting this in \eqref{Hint1} we get
\begin{equation}
\label{su2Hint}
H_{\rm int} = - \frac{1}{8\pi^2 N} \tr ( [ a_\uparrow^\dagger,a_\downarrow^\dagger][a^\uparrow,a^\downarrow] )
\end{equation}
The total Hamiltonian of $SU(2)$ Spin Matrix theory is then
\begin{equation}
\label{su2H}
H = L + g H_{\rm int} 
\end{equation}
where $L$ is given in \eqref{Lop}. In general one could also include a $- \mu S_z$ term in \eqref{su2H} but we choose $\mu=0$ in the following. The partition function that we analyze in the following is thus
\begin{equation}
\label{partfctsu2}
Z(\beta) = \tr ( e^{-\beta (L + g H_{\rm int}) })
\end{equation}

\subsection{Planar limit}
\label{sec:planarlimit}

Consider $SU(2)$ Spin Matrix theory in the planar limit $N=\infty$. As explained in Section \ref{sec:SCfromSMT} Spin Matrix theory is a gas of spin chains in the planar limit. For this specific Spin Matrix theory we have $U^{s'r'}_{sr}$ given by \eqref{U_su2} from which one infers the spin chain Hamiltonian acting on a spin chain as in \eqref{Hintspin}. This corresponds to the Hamiltonian for the ferromagnetic XXX$_{1/2}$ Heisenberg spin chain \cite{Minahan:2002ve}.%
\footnote{With an extra term $- \mu S_z$ in \eqref{su2H} one would get the ferromagnetic XXX$_{1/2}$ Heisenberg spin chain with a magnetic field \cite{Harmark:2006ie}.}
The partition function for $SU(2)$ Spin Matrix theory \eqref{partfctsu2} is obtained from the partition function of the Heisenberg spin chain as follows \cite{Harmark:2006ta}
\begin{equation}
\label{ZfromZXXX}
\log Z(\beta) = \sum_{n=1}^\infty \sum_{L=1}^\infty \frac{1}{n} e^{-\beta n L} Z_L^{\rm (XXX)} (n\beta)
\end{equation}
where $Z_L^{\rm (XXX)} = \tr_L ( e^{-\beta g H_{\rm int} })$ is the partition function of the Heisenberg spin chain theory with the trace $\tr_L$ being over spin chains (single-trace states) of length $L$. 

We see that the $SU(2)$ Spin Matrix theory partition function $Z(\beta)$ is that of a non-interacting gas of Heisenberg spin chains. Raising the temperature $T=1/\beta$ we encounter a singularity in the partition function at a temperature $T=T_{\rm H}(g)=1/\beta_{\rm H}(g)$ which is a function of the coupling constant $g$. This is the Hagedorn temperature for $SU(2)$ Spin Matrix theory. We call it a Hagedorn temperature since the density of states goes like $e^{\beta_{\rm H} E}$ for high energies. Defining the free energy per site of the Heisenberg spin chain
\begin{equation}
F(\beta) = - \frac{1}{\beta} \lim_{L \rightarrow \infty} \frac{1}{L} \log Z_L^{\rm (XXX)} (\beta)
\end{equation}
the Hagedorn temperature at any $g$ satisfies $F(\beta_H) = -1$ \cite{Harmark:2006ta}. One can thus find the Hagedorn temperature at any coupling from this. In particular for the weak and strong coupling regimes  \cite{Harmark:2006ta}
\begin{equation}
\label{su2TH}
T_H = \left\{ \begin{array}{l}\ds
 \frac{1}{\log 2} + \frac{g}{2^4 \pi^2 \log 2}  - \frac{3 g^2}{2^9 \pi^4}  +  \frac{(3+2\log2 )g^3}{2^{13} \pi^6} + \CO( g^4) \ \ \mbox{for} \ \
g \ll 1 
\\[4mm] \ds \frac{g^{\frac{1}{3}}}{(2\pi)^{\frac{1}{3}} \zeta ( \frac{3}{2} )^{\frac{2}{3}}} + \frac{4\pi}{3\, \zeta (\frac{3}{2} ) ^2} + \CO( g^{-\frac{1}{3}} )\ \ \mbox{for} \ \
g \gg 1
\end{array} \right.
\end{equation}

The resemblance between $SU(2)$ Spin Matrix theory and the AdS/CFT correspondence in the planar limit is evident in the sense that for finite coupling $g$ the theory is determined from an integrable spin chain (analog to the $psu(2,2|4)$ spin chain). Moreover, for weak coupling $g \ll 1$ the spectrum of $SU(2)$ Spin Matrix theory is that of the length operator $L$ plus a small perturbation from $g H_{\rm int}$. This corresponds to the spectrum of  weakly coupled planar $\CN=4$ SYM in the $SU(2)$ sector. For weak coupling $g\ll 1$ one can interpret the phase below the Hagedorn temperature as a confined phase, with the confinement arising from the singlet condition on the matrix indices of $a_s^\dagger$. 

For strong coupling $g \gg 1$ the spectrum of the Hamiltonian $H= L + gH_{\rm int}$ of $SU(2)$ Spin Matrix theory is in terms of states with $H_{\rm int} \ll 1$. For a single-trace state with $L \gg 1$ ($i.e.$ a spin chain) this corresponds to scattering magnons with low momenta of order $1/L$. The low energy spectrum is expanded in powers of $1/L$. To leading order $H_{\rm int} = \frac{1}{2L^2} \sum_{n\neq 0} n^2 M_n$ with $\sum_{n\neq 0} n M_n=0$ where $M_n$ is the number of particles with level number $n$. This resembles a quantum string spectrum. Note that the $n^2$ dependence signifies a Galilean dispersion relation $E \propto p^2$.
Employing coherent states one can go to a semi-classical regime of the spin chain and write down an effective action for large $L$. This gives the Landau-Lifshitz sigma-model action with target space $S^2$ \cite{Fradkin}. Thus, one can obtain both something resembling quantum strings as well as semi-classical strings with continous world-sheet and a geometric target space from the $SU(2)$ Spin Matrix theory at strong coupling $g \gg 1$. Note that while all these considerations are for single-trace states, the full spectrum of multi-trace states takes the significance of a free gas of strings. The Hagedorn temperature is thus the Hagedorn temperature of this gas of strings. 

The three phases with the confining phase of planar $\CN=4$ SYM in the $SU(2)$ sector for $g \ll 1$, a gas of Heisenberg spin chains for finite $g$, and a non-relativistic string theory for large $g$ are depicted as the three low temperature phases in the $(g,T)$ phase diagram of Figure \ref{fig:SMT}.

\subsubsection*{Match with string theory on $\ads_5\times S^5$}

In \cite{Harmark:2006ta,Harmark:2008gm} it was argued that one can match the planar $\CN=4$ SYM on $\R \times S^3$ in the $SU(2)$ Spin Matrix theory limit for $g\gg 1$ to string theory on $\ads_5\times S^5$ in the dual limit. This is a non-trivial claim in that $\lambda \rightarrow 0$ usually means going to a quantum string regime. However, in \cite{Harmark:2008gm} it is explained that the limit can be taken on the string theory side since: 1) One ends up with a semi-classical action (for large $L$). 2) The modes that decouple in the limit become infinitely heavy. 3) Thanks to supersymmetry the sigma-model that one starts with is robust to taking $\lambda\rightarrow 0$. 4) Zero-mode quantum fluctuations are suppressed since one is considering small fluctuations around a half-BPS state.

Indeed, one finds that in the limit the semi-classical sigma-model of tree-level string theory on $\ads_5 \times S^5$ reduces to the Landau-Lifshitz model mentioned above that one finds for $g\gg 1$ on the gauge theory side. This can explain the otherwise mysterious one-loop match between the gauge and string sides in the AdS/CFT correspondence \cite{Harmark:2008gm}.

Note in particular that one can take the pp-wave limit of \cite{Bertolini:2002nr} first and then afterwards the $SU(2)$ Spin Matrix theory limit.%
\footnote{In this pp-wave background the dispersion relation is $\sqrt{1+ \frac{\lambda p^2}{4\pi^2}}-1$ from which one sees very clearly that the $SU(2)$ Spin Matrix theory limit is a non-relativistic limit \cite{Harmark:2008gm}.}
 In this case one finds a direct match between the (limit of) the Hagedorn temperature computed in string theory on the pp-wave background and the Hagedorn temperature \eqref{su2TH} for $g \gg 1$. This provided the first match of the Hagedorn temperature in the AdS/CFT correspondence \cite{Harmark:2006ta}.

%%%%%%%%%%%%%%%%%%%%%%%%%%%%%%%%%%%%%%%%%%%%%%%%%%%%%%%%%%
\subsection{Large $N$ and low temperature: Nearly-planar regime}
\label{sec:nearlyplanar}

For large but finite $N$ the non-planar corrections to the planar limit are small for low temperatures or low energies. A way to see this is to start in the planar limit $N=\infty$. Then the expectation value of $L$ is finite for $T < T_{\rm H}(g)$ but diverges for $T \rightarrow T_{\rm H}(g)$. Hence reintroducing a large but finite $N$ the expectation value of $L$ reaches $N$ at a temperature $T_{\rm mix}(g,N) < T_{\rm H}(g)$ and above this temperature the theory becomes increasingly non-planar since single-trace states start to mix with multi-trace states. Conversely, for low temperatures $T<T_{\rm mix}(g,N)$ the planar limit is a good approximation for large $N$. Hence one can think of the planar limit as a low temperature (or low energy) limit.

While the planar limit is a good approximation for $T<T_{\rm mix}(g,N)$ one has corrections starting at order $1/N$. Considering the action of $H_{\rm int}$ on a contracted two oscillator state Eq.~\eqref{Hinttwoosc} one sees that while the second term on the right hand side keeps the same matrix index structure the three other terms, which are of order $1/N$, either move the contraction to involve only one oscillator or make it a double contraction. More generally, one can infer from Eqs.~\eqref{Hint1} and \eqref{sigmaT} that acting with $H_{\rm int}$ on a multi-trace state of the form \eqref{tracebasis} one gets states with the same matrix contractions (but different spin indices) to zeroth order in $1/N$. Subleading to this are terms of order $1/N$ where either one of the single-traces are broken up in two, or two of the single-traces are joined into one (with different spin indices) \cite{Bellucci:2004ru,Peeters:2004pt,Casteill:2007td}.
Since to leading order the single-traces can be interpreted as individual spin chains, the subleading $1/N$ terms can be interpreted as describing the splitting or joining of spin chains, thus providing an interaction between the spin chains. Therefore, when going from $N=\infty$ to large but finite $N$ the free gas of spin chains becomes a weakly interacting gas of spin chains.%
\footnote{Note that the so-called spin bit model of \cite{Bellucci:2004ru} is based on the same interaction as $H_{\rm int}$ for the $SU(2)$ Spin Matrix theory. This work focusses on the perturbative $1/N$ effects, and the interaction is studied in terms of splitting and joining effects of spin chains. Note also the related work on bit strings of \cite{Vaman:2002ka} which considers $1/N$ effects from the string theory point of view.}

For temperatures sufficiently above $T_{\rm mix}(g,N)$ one encounters a phase transition where the planar limit does not anymore provide the leading large $N$ behavior. At this phase transition we go from a confining behavior $\log Z \sim \CO(1)$ to a deconfining behavior $\log Z \sim \CO(N^2)$ of the partition function $Z$. While for $g=0$ one finds that this phase transition occurs at the Hagedorn temperature $T_H = 1/\log 2$, for $g>0$ the phase transition to deconfining behavior might very well occur below $T_H(g)$. In Section \ref{sec:hightemp} we investigate the phases of $SU(2)$ Spin Matrix theory in the large temperature, deconfining regime.

%%%%%%%%%%%%%%%%%%%%%%%%%%%%%%%%%%%%%%%%%%%%%%%%%%%%%%%%%%
\section{$SU(2)$ Spin Matrix theory at high temperature (non-planar)}
\label{sec:hightemp}

In this section we explore the phases of $SU(2)$ Spin Matrix theory, as illustrated in Figure \ref{fig:SMT}, for high temperatures. In Sections \ref{sec:suq_partfcts} and \ref{sec:classical_phase_zero} we study in detail the partition function of free $SU(q)$ Spin Matrix theory, showing in particular that it corresponds to $(q-1)N^2+1$ harmonic oscillators at high temperatures. In Section \ref{sec:clas_any} we find a classical matrix model that describes $SU(2)$ Spin Matrix theory at any coupling in the high temperature limit and we show that for $g\rightarrow \infty$ one obtains a theory of $2N$ harmonic oscillators.

%%%%%%%%%%%%%%%%%%%%%%%%%%%%%%%%%%%%%%%%%%%%%%%%%%%%%%%%%%
\subsection{Free $SU(q)$ Spin Matrix theory at high temperature}
\label{sec:suq_partfcts}

For $g=0$ one can compute the partition function \eqref{theZ} for Spin Matrix theory exactly with any representation $R_s$ for the spin indices by employing the techniques of \cite{Sundborg:1999ue,Aharony:2003sx}. To compute this one needs the partition function for a single spin index
\begin{equation}
z(\beta,\mu_p) = \sum_{s\in R_s} \langle s | e^{-\beta  + \sum_p \beta \mu_p K_p}  |s \rangle 
\end{equation}
The $g=0$ partition function is then
\begin{equation}
\label{freepartfct1}
Z(\beta,\mu_p)|_{g=0} = \int [dU] \exp \left( \sum_{n=1}^\infty \frac{z(n\beta,\mu_p)}{n}  \tr U^n \tr (U^\dagger)^n \right)
\end{equation} 
where $U \in U(N)$ is a $N \times N$ unitary matrix and $[dU]$ is the integration over the unitary matrices with the Haar measure. We assumed for simplicity that $R_s$ is bosonic (when including fermionic states one should put an extra minus sign for even $n$ when computing $z(n\beta,\mu_p)$).
Using Frobenius' formula one can write Eq.~\eqref{freepartfct1} in terms of characters of the symmetric group \cite{Dutta:2007ws}
\begin{equation}
\label{freepartfct2}
Z(\beta,\mu_p)|_{g=0}=1+\sum_{n=1}^\infty \sum_k \sum_r \prod_{j=1}^n \frac{z(j\beta,\mu_p)^{k_j}}{k_j! j^{k_j}} | \chi(r,k) |^2
\end{equation}
Here $n$ labels the symmetric groups $S_n$ and $k=(k_1,....,k_n)$ labels the conjugacy classes of $S_n$ with $k_j$ being the number of $j$-cycles. Moreover, $r=[r_1,...,r_n]$ labels the irreducible representations of $S_n$ which can be represented as Young tableaux $[r_1,...,r_n]$ with $n$ boxes and at most $N$ rows since these are in correspondence to representations of $U(N)$. Here $r_j$ in $[r_1,...,r_n]$ is the number of boxes in the $j$'th row. Finally, $\chi(r,k)$ is the character for the symmetric group $S_n$ for the representation $r$ and conjugacy class $k$ (see for example \cite{Hamermesh} for computations of $\chi(r,k)$). 

We apply now the general formula \eqref{freepartfct2} to free $SU(q)$ Spin Matrix theory ($R_s$ being the fundamental representation). We take the special case $\mu_p=0$. Thus,
\begin{equation}
\label{smallz}
z(n\beta) = q x^n \spa x \equiv e^{-\beta}
\end{equation}
The partition function \eqref{freepartfct2} reduces to%
\footnote{Note that this partition function counts the number of restricted Schur polynomials  \cite{Balasubramanian:2004nb,Koch:2007uu} of $q$ variables weighted by their lengths. This follows from the fact that the restricted Schur polynomials provides a basis for $SU(q)$ Spin Matrix theory as noted in Section \ref{sec:defSMT}.} 
\begin{equation}
\label{freepartfct3}
Z_{q,N}(\beta)|_{g=0}=1+\sum_{n=1}^\infty x^n \sum_k \sum_r \prod_{j=1}^n \frac{q^{k_j}}{k_j! j^{k_j}} | \chi(r,k) |^2
\end{equation}
Here we added the extra indices $q$ and $N$ to highlight the dependence on these parameters.
The goal in the following is to understand the behavior of this partition function for large temperature, $i.e.$ $x \rightarrow 1$, given $q$ and $N$.

For $q=1$ the partition function \eqref{freepartfct3} corresponds to $U(1)$ Spin Matrix theory (this has $H_{\rm int}=0$ since $R_s$ is one-dimensional). In this case one finds using $\sum_k \frac{1}{k_j! j^{k_j}} | \chi(r,k) |^2 = 1$ that $Z_{1,N}(\beta)|_{g=0} = 1 + \sum_{n=1}^\infty C_{n,N} x^n$ where $C_{n,N}$ is the number of Young tableaux with $n$ boxes and at most $N$ rows. From this one finds
\begin{equation}
Z_{1,N} (\beta)|_{g=0} = \prod_{n=1}^N \frac{1}{1-x^n}
\end{equation}
In this case there is no Hagedorn singularity of the partition function for $N=\infty$. Hence the planar limit of the partition function is valid for $T \ll N$. For $T \gg N$ one has $1-x^n \simeq n \beta$ for $n=1,2,...,N$ and hence
$Z_{1,N} (\beta)|_{g=0} \simeq \frac{1}{N!} (1-x)^{-N}$ which is the partition function for $N$ indistinguishable one-dimensional harmonic oscillators. For large $N$ we have $\log Z_{1,N} (\beta)|_{g=0} \simeq - N \log N + N \log T$. We see that in fact the $N\log N$ term can be neglected for $T \gg N$ and hence $\log Z_{1,N} (\beta)|_{g=0} \simeq   N \log T$ which is the partition function for $N$ distinguishable one-dimensional harmonic oscillators. Indeed, it is a general fact in statistical physics that both the Bose-Einstein statistics and the Fermi-Dirac statistics for indistinguishable particles asymptote for high temperatures to the Maxwell-Boltzmann statistics in which all particles are distinguishable (this is known in statistical physics as the {\sl classical limit}). Thus, we can conclude that the high temperature phase of $U(1)$ Spin Matrix theory is $N$ one-dimensional harmonic oscillators.

For $q \geq 2$ a general formula that resums the infinite series over $n$ in Eq.~\eqref{freepartfct3} is not known. Only in the special case of $N=\infty$ one finds%
\footnote{Using that the sum over $r$ is unrestricted one has $\sum_r |\chi(r,k)|^2 = \prod_{j=1}^n k_j ! j^{k_j}$. Then one can see that $Z_{q,N=\infty}(\beta)|_{g=0} = 1 + \sum_{n=1}^\infty q^n (Z_{1,n}(\beta)|_{g=0} - Z_{1,n-1}(\beta)|_{g=0} )$ which gives Eq.~\eqref{planarZqN}.}
\begin{equation}
\label{planarZqN}
Z_{q,N=\infty}(\beta)|_{g=0} = \prod_{n=1}^\infty \frac{1}{1-qx^n}
\end{equation}
Therefore, we have devised a method to compute $Z_{q,N}(\beta)|_{g=0}$ for particular values of $q$ and $N$. We assume that $Z_{q,N}(\beta)|_{g=0}$ is of the form $P(x)/Q(x)$ where $P(x)$ and $Q(x)$ are two polynomials. 
Given this assumption one can compute $Z_{q,N}(\beta)|_{g=0}$ by computing a finite number of coefficents of $x^n$ in Eq.~\eqref{freepartfct3} (one can obviously test the assumption by computing extra terms as well). In Appendix \ref{app:partfcts} we listed the results for a number of different values of $q \geq 2$ and $N \geq 2$.
It is not clear from the obtained partition functions what the general form is. Nevertheless, for the high temperature limit $x\rightarrow 1$ we find that all the partition functions are of the form
\begin{equation}
\label{largeTpre}
Z_{q,N} (\beta)|_{g=0} \simeq \frac{a_{q,N}}{(1-x)^{(q-1)N^2+1}} \ \ \mbox{for} \ \ T \rightarrow \infty
\end{equation}
We conjecture that this is the high temperature form for all $q\geq2$ and $N\geq 2$.%
\footnote{We listed the coefficients $a_{q,N}$ in Table \ref{tab:aqN} in Appendix \ref{app:partfcts}. These coefficents do not provide any obvious interpretation in terms of the statistics of indistinguishable particles.}
As a complement to the analysis of exact partition functions for finite $N$ we study in Appendix \ref{app:eigenval} the partition function \eqref{freepartfct3} numerically in the large $N$ limit using the technique of integrating over the eigenvalues of a unitary matrix. We find for $q=2,3,4,5$
\begin{equation}
\label{largeTlargeN}
\lim_{N\rightarrow \infty} \frac{1}{N^2} \log Z_{q,N} (\beta)|_{g=0} \simeq - \log b_q + (q-1) \log T \ \ \mbox{for} \ \ T \gg \frac{1}{\log q}
\end{equation}
where $b_q$ is a constant (for all $q$ we find $b_q\simeq 8.9$). This is in accordance with the conjecture \eqref{largeTpre}. Since $b_q$ is a number of order one, one can also infer that the classical limit, where one obtains Maxwell-Boltzmann statistics, consists in having $T \gg 1$ even for large $N$. In conclusion, in the classical limit $T \gg 1$ we can neglect the coefficient $a_{q,N}$ of \eqref{largeTpre} (or $b_q$ in \eqref{largeTlargeN}) and hence one finds
\begin{equation}
\label{largeTconjecture}
\log Z_{q,N} (\beta)|_{g=0} \simeq  [ (q-1) N^2+1 ] \log T \ \ \mbox{for} \ \ T \gg 1
\end{equation}
which one recognizes as the partition function of $(q-1)N^2 +1$ one-dimensional harmonic oscillators.

According to \eqref{largeTconjecture} the high-temperature phase of free $SU(q)$ Spin Matrix theory thus corresponds to $(q-1)N^2+1$ harmonic oscillators. Hence large $N$ free $SU(q)$ Spin Matrix theory exhibits a phase transition at the Hagedorn temperature $T_{\rm H}=1/\log q$ into what we call a partially deconfined phase since it bears resemblance to having full deconfinement, in particular with the feature that the coupling between the $(q-1)N^2+1$ harmonic oscillators goes to zero as $T\rightarrow \infty$. Full deconfinement would mean $q N^2$ uncoupled one-dimensional harmonic oscillators at high temperatures since this is what $SU(q)$ Spin Matrix theory would correspond to without the singlet condition ($i.e.$ using Hilbert space $\CH'$ instead of $\CH$ in Eqs.~\eqref{thebasisprime} and \eqref{tracebasis}). However, while we do get uncoupled harmonic oscillators at high temperatures, we get $N^2-1$ less than what one would have with full deconfinement. Below in Section \ref{sec:classical_phase_zero} we shall see how the number $N^2-1$ emerges from the singlet condition.

Note that the high-temperature phase for $N \rightarrow \infty$ goes like $(q-1)N^2\log T$. Instead the confined phase below the Hagedorn temperature is of order one with respect to $N$. Thus, using $F/N^2$ as an order parameter, with $F=-T\log Z$ being the free energy, we see that we exhibit a phase transition from the confined phase at low temperature, with $F/N^2=0$, to the partially deconfined phase at high temperature, with $F/N^2 = - (q-1) T \log T$.

It is interesting to search for an interpretation of the partial deconfinement of the spin chain gas. A possible explanation could be that the spin chains breaks up into smaller independent constituents that previously were bound together in the confined phase. In part this is true since it is known that certain single-trace configurations with lengths exceeding $N$ can be split up into combinations of shorter single-traces. However, this explanation is flawed. In Appendix \ref{app:partfcts} we consider the Plethystic logarithm \cite{Benvenuti:2006qr} of the obtained partition functions. The Plethystic logarithm gives back the single-trace partition function that can generate the full multi-trace partition function. Thus, if there were just a few single-traces that could generate the full Hilbert-space the result of the Plethystic logarithm should be a polynomial of low degree. This is indeed true for the cases for $(q,N)=(2,2), (2,3), (3,2)$.
However, these cases seems to be the exceptions to the rule. For any case with higher $q$ or $N$ than in these three cases the result is an infinite series in $x$. As an example consider $(q,N)=(4,2)$ for which the Plethystic logarithm gives an infinite series. Even if this is merely 2 by 2 matrices with four different spin labels one has to include single-traces and algebraic relations between them with arbitrarily large lengths. This is obviously in contrast with the fact that one obtains a relatively simple theory at high temperatures with just 13 one-dimensional harmonic oscillators. Thus, we do not have any clear identification between the single-trace or multi-trace states and the raising operators for the $(q-1)N^2+1$ harmonic oscillators. 

In conclusion, looking at the partition functions for free $SU(q)$ Spin Matrix theory, it seems clear that our best hope for a regime in which one can obtain a systematic understanding of the large $N$ non-planar behavior of $SU(2)$ Spin Matrix theory is in the high temperature classical limit. Indeed, we achieve this below by finding a classical description of $SU(2)$ Spin Matrix theory, even for arbitrary coupling $g$.

%%%%%%%%%%%%%%%%%%%%%%%%%%%%%%%%%%%%%%%%%%%%%%%%%%%%%%%%%%
\subsection{Classical description of high-temperature regime}
\label{sec:classical_phase_zero}

Above we found that in the classical limit of the partition functions of free $SU(q)$ Spin Matrix theory one gets the partition function of $(q-1)N^2+1$ one-dimensional harmonic oscillators. The classical limit in thermodynamics means that we have such a highly excited system that the quantum statistical mechanics is well approximated by classical statistical mechanics. Thus, one should be able to find a classical description of the thermodynamics in this regime, $e.g.$ where one can obtain the partition function by integrating over the classical phase space.
We use here the method of coherent states to find the description of free $SU(q)$ Spin Matrix theory in the classical limit. In Section \ref{sec:clas_any} we generalize this description to any coupling $g$ in the case of $SU(2)$ Spin Matrix theory.

\subsubsection*{Coherent state description}

The coherent states of the Hilbert space $\CH$ spanned by \eqref{tracebasis} for free $SU(q)$ Spin Matrix theory are given as
\begin{equation}
\label{coh_state}
|\lambda \rangle = \CN_\lambda \exp \Big( \sum_s \tr ( \lambda_s a^\dagger_s  ) \Big) |0\rangle \spa \langle \lambda | \lambda \rangle = 1
\end{equation}
where $\lambda_s$, $s=1,...,q$, are $q$ complex $N \times N$ matrices with entries $(\lambda_s)^i{}_j$ that specify the coherent state. We split them up in Hermitian and anti-Hermitian parts
\begin{equation}
\lambda_s = \frac{1}{\sqrt{2}} ( X_s + i P_s ) \spa s = 1, ..., q
\end{equation}
where $X_s$ and $P_s$ are Hermitian $N\times N$ matrices.
The coherent state \eqref{coh_state} has the properties
\begin{equation}
\label{coh_props}
(a^s)^i {}_j |\lambda \rangle = (\lambda_s)^i {}_j |\lambda \rangle \spa \langle \lambda | (a_s^\dagger)^i {}_j  = \langle \lambda | (\lambda_s^\dagger)^i {}_j 
\end{equation}
As such \eqref{coh_state} is a state in $\CH'$. To make it into a state of $\CH$ we should impose the singlet condition \eqref{singletcond}. Since we are in a (semi-)classical regime it is enough to demand that the expectation value of the operator $\Phi^i{}_j$ is zero, giving
\begin{equation}
0  = \langle \lambda | \Phi^i{}_j |\lambda \rangle = \Big( \sum_s [ \lambda_s^\dagger, \lambda_s] \Big)^i{}_j = \Big( i \sum_s [ X_s,P_s] \Big)^i{}_j
\end{equation}
which we see amounts to imposing Gauss constraint 
\begin{equation}
\label{suqgauss}
\sum_s [ X_s , P_s] = 0
\end{equation}
Thus by imposing the Gauss constraint on $\lambda_s$ the coherent state \eqref{coh_state} is a state in $\CH$. Turning to the Hamiltonian $H = L $  we compute the classical Hamiltonian $H_{\rm cl}(X_s,P_s)$
\begin{equation}
\label{clashamsuq}
H_{\rm cl}(X_s,P_s) = \langle \lambda | H |\lambda \rangle = \langle \lambda | \sum_s \tr ( a_s^\dagger a^s  ) |\lambda \rangle  = \sum_s \tr ( \lambda_s^\dagger \lambda_s) = \frac{1}{2} \sum_s \tr ( P_s^2 + X_s^2 )
\end{equation}
We notice that in the absence of the Gauss constraint \eqref{suqgauss} the Hamiltonian \eqref{clashamsuq} describes a system of $qN^2$ uncoupled one-dimensional harmonic oscillators. This is the classical analog of the statement that without the singlet condition free $SU(q)$ Spin Matrix theory would describe $qN^2$ uncoupled quantum harmonic oscillators.%
\footnote{The Lagrangian that corresponds to the Hamiltonian \eqref{clashamsuq} and constraint \eqref{suqgauss} is $L = \frac{1}{2} \tr \left( \sum_s \Big[ (D_0 X_s)^2 - X_s^2 \Big] \right)$ with $D_0 X_s = \dot{X}_s + i [A_0,X_s]$ where the Hermitian matrix $A_0(t)$ is a gauge field. We can choose the gauge $A_0(t)=0$ in which case we get the Lagrangian and Gauss constraint
\begin{equation}
\label{LGfree}
L = \frac{1}{2} \tr \left( \sum_s \Big[ \dot{X}_s^2 - X_s^2 \Big] \right) \spa \sum_q [X_s,\dot{X}_s]=0
\end{equation}
Note that the Gauss constraint is a non-holonomic constraint of a type that one can deal with by introducing fictitious forces when deriving the equations of motion \cite{Goldstein_2nd_edition}. However, for this particular constrained theory these fictitious forces are zero due to gauge invariance of the theory without gauge fixing and hence the equations of motion are simply $X_s + \ddot{X}_s=0$.}

The classical partition function is
\begin{equation}
\label{clasZ}
Z_{q,N}(\beta)|_{g=0} = \frac{1}{(2\pi)^{(q-1)N^2+1}} \int dP dX e^{-\beta H_{\rm cl}(X_s,P_s)} \delta \big( C(X_s,P_s) \big)
\end{equation}
where we defined $C(X_s,P_s) \equiv \sum_s [X_s,P_s]$. 
In the partition function \eqref{clasZ} we are approximating $\tr ( e^{-\beta H} )$ by integrating $e^{-\beta H_{\rm cl}}$  over the classical phase space while imposing the Gauss constraint.
This partition function is a good approximation to the exact partition function \eqref{freepartfct3} at high temperatures $T \gg 1$ (the classical limit).

We can now count the number of independent constraints included in the Gauss constraint in \eqref{suqgauss}. Since the left-hand side of the constraint is an anti-Hermitian and traceless matrix it has $N^2-1$ independent real parameters. Hence, we propose that at high temperatures in the above classical description the $N^2-1$ real constraints from the Gauss constraint are responsible of the fact that we have $N^2-1$ less harmonic oscillators than if one did not impose the Gauss constraint, thus providing an explanation for having $(q-1)N^2+1$ one-dimensional harmonic oscillators at high temperatures in free $SU(q)$ Spin Matrix theory. While this seems clear at the level of counting constraints and oscillators, in practise the $N^2-1$ constraints are difficult to solve in general. In other words, the above classical constrained Hamiltonian system does not correspond to uncoupled harmonic oscillators  at finite temperatures, the uncoupled harmonic oscillators emerge only at high temperatures. Below we give an example for $(q,N)=(2,2)$.%
\footnote{For certain matrix models with a single complex matrix $Z$ one can bring $Z$ to the form $Z=U T U^\dagger$ using an $SU(N)$ transformation $U$ such that $T$ is an upper diagonal matrix with $N^2+1$ real parameters (see for example \cite{Eynard:2005wg}) and subsequently show that the dependence on $U$ drops out of the theory. However, it does not apply to the above case for $q=2$ and $Z=X_1+iX_2$ with Lagrangian and Gauss constraint \eqref{LGfree} because of the kinetic term for $Z$. To see this take $Z=T$ thus with Gauss constraint $[T,\dot{T}^\dagger] + [T^\dagger,\dot{T}]=0$ which is not satisfied in general for an upper-triangular matrix $T$. For $N=2$ one can check this explicitly for $T= \matrto{z_1}{m}{0}{z_2}$ with $m\in \R$ and $z_1,z_2\in \C$.}

\subsubsection*{Check of emerging uncoupled harmonic oscillators at high temperature}

We now make an explicit check for $(q,N)=(2,2)$ to see that one obtains a classical partition function corresponding to five uncoupled one-dimensional harmonic oscillators. Write
\begin{equation}
X_1 = \frac{1}{\sqrt{2}} \matrto{x_1+x_2}{x_3+ix_4}{x_3-ix_4}{x_1-x_2} \spa 
X_2 = \frac{1}{\sqrt{2}} \matrto{x_5+x_6}{x_7+ix_8}{x_7-ix_8}{x_5-x_6}
\end{equation}
The equations of motion are $\ddot{x}_i + x_i = 0$, $i=1,2,...,8$, and the three Gauss constraints are $J_{23} + J_{67} = 0$, $J_{24} + J_{68} = 0$ and $J_{34} + J_{78} = 0$ where we defined the angular momenta $J_{ij} \equiv x_i \dot{x}_j - x_j \dot{x}_i$. Note that for any solution to the equations of motion one has $\frac{d}{dt} J_{ij} = 0$ thus the angular momenta are constants of motion. 

We can satisfy the two constraints $J_{24} + J_{68} = 0$ and $J_{34} + J_{78} = 0$ by setting $x_4 = 0$ and $x_8=0$ (which are also two constraints). Doing this, we still need to impose $J_{23} + J_{67} = 0$. We make the parametrization $x_1 = x$, $x_2 = r \cos \phi$, $x_3= r\sin \phi$, $x_5=y$, $x_6= l \cos \varphi$ and $x_7=l \sin \varphi$. The Hamiltonian \eqref{clashamsuq} is
\begin{equation}
H_{\rm cl} = \frac{1}{2} \left( p_x^2 + x^2 + p_y^2 + y^2+  p_r^2 + \frac{p_\phi^2}{r^2} + r^2 + p_l^2 + \frac{p_\varphi^2}{l^2} + l^2\right)
\end{equation}
The constraint is $p_\phi + p_\varphi = 0$. The classical partition function is
\begin{equation}
Z_{2,2}(\beta)|_{g=0} = \frac{1}{(2\pi)^5} \int dx dy dr dl d\phi d\varphi dp_x dp_y dp_r dp_l dp_\phi dp_\varphi e^{- \beta H } \delta ( p_\phi + p_\varphi )
\end{equation}
We compute
\begin{eqnarray}
Z_{2,2}(\beta)|_{g=0} &=&    \frac{1}{ \beta^3} \int dr dl dp_\phi e^{-\frac{1}{2} \beta (  \frac{p_\phi^2}{r^2} + r^2 + \frac{p_\phi^2}{l^2} + l^2) } = \frac{1}{ \beta^3} \int dp_\phi \left( \sqrt{\frac{\pi}{2\beta}} e^{-\beta |p_\phi|} \right)^2
\nn \\ &=&     \frac{\pi}{\beta^4} \int_0^\infty dp_\phi  e^{-2\beta p_\phi} = \frac{\pi}{2\beta^5}
\end{eqnarray}
Thus, since we are at high temperatures $T \gg 1$ we find $\log Z_{2,2}(\beta)|_{g=0} \simeq - 5 \log T$ which indeed is the partition function of five uncoupled one-dimensional harmonic oscillators.

%%%%%%%%%%%%%%%%%%%%%%%%%%%%%%%%%%%%%%%%%%%%%%%%%%%%%%%%%%
\subsection{Classical matrix model for $SU(2)$ Spin Matrix theory at any coupling}
\label{sec:clas_any}

In this section we use the coherent state method to find the classical Hamiltonian for $SU(2)$ Spin Matrix theory at any coupling $g$. This provides a classical description of $SU(2)$ Spin Matrix theory for sufficiently high temperatures (the classical limit).

The coherent states of the Hilbert space $\CH$ for $SU(2)$ Spin Matrix theory are given by
\begin{equation}
\label{coh_state2}
\begin{array}{c}\ds
|\lambda \rangle = \CN_\lambda \exp \Big( \tr ( \lambda_\uparrow a^\dagger_\uparrow + \lambda_\downarrow a^\dagger_\downarrow  ) \Big) |0\rangle \spa \langle \lambda | \lambda \rangle = 1
\\[4mm] \ds
\lambda_s = \frac{1}{\sqrt{2}} ( X_s + i P_s ) \spa s = \uparrow,\downarrow
\end{array}
\end{equation}
with the Gauss constraint
\begin{equation}
\label{su2gauss}
[ X_\uparrow , P_\uparrow] + [X_\downarrow,P_\downarrow] = 0
\end{equation}
that follows from the expectation value of the singlet condition $\langle \lambda | \Phi^i{}_j |\lambda \rangle=0$. These coherent states are a special case of the ones of Section \ref{sec:classical_phase_zero}. They have properties \eqref{coh_props} for $s=\uparrow,\downarrow$. Turning to the Hamiltonian $H = L + g H_{\rm int}$ we have already computed the free part in $\langle \lambda | L | \lambda \rangle$ in \eqref{clashamsuq}. For the interacting part we have
\begin{equation}
\begin{array}{l}\ds
\langle \lambda | H_{\rm int} |\lambda \rangle = - \frac{1}{8\pi^2 N} \tr ( [\lambda_\uparrow^\dagger,\lambda_\downarrow^\dagger][\lambda_\uparrow,\lambda_\downarrow])  \\[2mm] \ds = - \frac{1}{32\pi^2 N} \tr \Big( [X_\uparrow,X_\downarrow]^2 + [P_\uparrow,P_\downarrow]^2 + [P_\uparrow,X_\downarrow]^2 + [X_\uparrow,P_\downarrow]^2 - 2 [X_\uparrow,P_\uparrow][X_\downarrow,P_\downarrow] \Big)
\end{array}
\end{equation}
Using now the Gauss constraint \eqref{su2gauss} on the last term the classical Hamiltonian $H_{\rm cl} (X_s,P_s)$ becomes
\begin{equation}
\label{final_clas_ham}
\begin{array}{l}\ds
H_{\rm cl} (X_s,P_s)=\langle \lambda | H |\lambda \rangle   = \frac{1}{2} \sum_s \tr ( P_s^2 + X_s^2 )   \\[2mm] \ds - \frac{g}{32\pi^2 N} \tr \Big( [X_\uparrow,X_\downarrow]^2 + [P_\uparrow,P_\downarrow]^2 + [X_\downarrow,P_\uparrow]^2 + [X_\uparrow,P_\downarrow]^2 +  [X_\uparrow,P_\uparrow]^2+  [X_\downarrow,P_\downarrow]^2 \Big)
\end{array}
\end{equation}
This Hamiltonian, together with the Gauss constraint \eqref{su2gauss}, describes the classical limit of the high-temperature regime of $SU(2)$ Spin Matrix theory at any coupling $g$. We call this a {\sl classical matrix model} since it is a classical constrained Hamiltonian system based on four $N\times N$ Hermitian matrices that provide an accurate description of the high temperature physics in the classical limit of the partition function. One can compute the classical partition function as we did in \eqref{clasZ}. For $g=0$ we know from the above that the classical matrix model corresponds to $N^2+1$ non-interacting harmonic oscillators. As $g$ is turned on, the classical matrix model, given Eqs.~\eqref{final_clas_ham} and \eqref{su2gauss}, describes an interacting system of $N^2 +1$ harmonic oscillators. 

It is interesting to consider the limit of large coupling $g \gg 1$ since this should resemble a string theory phase. Indeed, if we take $g\rightarrow \infty$ we see from the Hamiltonian that all the four matrices $X_\uparrow$, $P_\uparrow$, $X_\downarrow$ and $P_\downarrow$ should commute with each other. This can only be true in general if they all are diagonal. Thus, we end up with the Hamiltonian
\begin{equation}
H_{\rm cl}(X_s,P_s)\Big|_{g\rightarrow \infty} = \frac{1}{2} \sum_s \sum_{i=1}^N \Big[ (P_s)^i {}_i + (X_s)^i {}_i \Big]
 \end{equation}
which describes $2N$ uncoupled one-dimensional harmonic oscillators. Note that the Gauss constraint \eqref{su2gauss} is solved by having the four matrices diagonal.

In the phase diagram for $SU(2)$ Spin Matrix theory in Figure \ref{fig:SMT} we have depicted the high-temperature phases for all values of $g$, in addition to the low temperature phases reviewed in Section \ref{sec:lowT}. We note that as a concrete result of our analysis of $SU(2)$ Spin Matrix theory we have found out that whereas the connecting link between weak and strong coupling $g$ in the planar limit $N=\infty$, $e.g.$ for low temperatures, is a spin chain theory, for high temperature it is instead a classical matrix model. Thus, whereas for low temperatures it is the representation of the spin group $SU(2)$ that defines the theory it is the matrix representation of the $U(N)$ group that does it at high temperatures. This is a manifestation of how the finite-$N$ effects change the nature of the theory. 

Considering the large coupling regime $g \gg 1$ of the phase diagram of Figure~\ref{fig:SMT} we have a non-relativistic string theory (with small string coupling) at low temperatures and a theory of $2N$ harmonic oscillators at high temperatures. We find that this is a very interesting result since it is the first time one has been able to determine precisely what type of degrees of freedom emerges at high temperatures when one warms up a gas of strings to the point where it undergoes a phase transition. 
Normally, such a study would not be possible to perform since it would require a quantitative understanding of non-perturbative closed string theory. Our results relate to the long-standing discussion on the Hagedorn temperature of string theory, e.g. in the classic paper of Atick and Witten they argue that the Hagedorn transition should be analogous to the deconfinement transition of gauge theory \cite{Atick:1988si}. Here we see that the high-temperature phase indeed exhibits deconfinement, and in a sense also asymptotic freedom in that the $2N$ harmonic oscillators are uncoupled for sufficiently high temperatures. In view of all this it would be highly interesting to examine more closely the phase transition at $T_c$. 

We speculate that one can give the following physical interpretation of our results in the $g\gg 1$ regime: For low temperatures the strings interact weakly and live on a sphere $S^2$ (as one can see from the Landau-Lifshitz model). As one raises the temperature the effective interaction between the strings becomes stronger since they are more likely to meet. At the temperature $T_c$ one has a phase transition to a phase of $N$ particles. For sufficiently high temperatures these $N$ particles live on a plane $\R^2$ since they are so energetic that the $S^2$ effectively looks like $\R^2$. Indeed they correspond to $N$ two-dimensional harmonic oscillators.

Finally, we turn to the question of how to interpret our $g \gg 1$ result on the string theory side of the AdS/CFT correspondence. 
Since for $g \rightarrow \infty$ we are considering the states for which $H_{\rm int}=0$ this corresponds to states with $H = L = R_1 + R_2$ where $R_i$ are the three R-charges of $\CN=4$ SYM or the three angular momenta on $S^5$ on the string theory side. These states are $1/4$ BPS states (in accordance with Section \ref{sec:g_infty}). The partition function of such $1/4$ BPS states has previously been studied in \cite{Kinney:2005ej} where it was found that for sufficiently high energies or temperatures it corresponds to $N$ two-dimensional harmonic oscillators which in fact is the same as $2N$ one-dimensional harmonic oscillators in the classical limit. Hence for $g \rightarrow \infty$ we should be able to match our result with $1/4$ BPS states on the string theory side at large energies or temperatures. It seems natural to interpret the $2N$ harmonic oscillators in terms of a gas of $1/4$ BPS Giant Gravitons at high temperatures.

It would be highly interesting to clarify how one could possibly take the $SU(2)$ Spin Matrix theory limit on the string side. In \cite{Harmark:2008gm} this limit is taken for the sigma-model of type IIB string theory on $\ads_5\times S^5$ and one finds that it matches the $g \gg 1$ results for the $N=\infty$ case of $SU(2)$ Spin Matrix theory, as mentioned in Section \ref{sec:planarlimit}. For the $SU(2)$ Spin Matrix theory in the high temperature classical regime it would be natural to imagine that one should be able to get the Hamiltonian \eqref{final_clas_ham} from a limit of a classical action for objects that are non-perturbative in the string coupling $g_s$, such as D-branes.

%%%%%%%%%%%%%%%%%%%%%%%%%%%%%%%%%%%%%%%%%%%%%%%%%%%%%%%%%%
\section{Discussion and outlook}
\label{sec:discuss}

We first make some general remarks on the relation of Spin Matrix theory to the AdS/CFT correspondence, and subsequently list some of the open problems that would be interesting to pursue.

\subsubsection*{General remarks}

One can take two roads to Spin Matrix theory. One can study it as a quantum mechanical theory in its own right, being a finite-$N$ generalization of nearest-neighbor spin chain theories, or possibly with the motivation that it shares many features with the AdS/CFT correspondence while still being simple enough to solve, and that studying it can lead to general observations about finite-$N$ effects and the strong coupling limit. This is in the same spirit as for example Berenstein's toy model for the AdS/CFT correspondence \cite{Berenstein:2004kk}, which in our language is $U(1)$ Spin Matrix theory, or certain matrix models which have been employed to mimic dynamics of quantum black holes \cite{Festuccia:2006sa}. %,Iizuka:2008hg}. 
Berenstein's toy model of \cite{Berenstein:2004kk} has also been used to consider features of quantum black holes \cite{Balasubramanian:2005mg}. This model is non-interacting, thus one can regard $SU(2)$ Spin Matrix theory as a step towards a more accurate model with a non-trivial coupling constant and a phase transition from confining to deconfining behavior. 

The other road to Spin Matrix theory is to employ it as a connecting link to the string theory side of the AdS/CFT correspondence. We now discuss the prospects of the latter. 

We briefly discussed in the Introduction our general philosophy, illustrated in Figure \ref{fig:philosophy}, about using Spin Matrix theory limits to get an improved understanding of the AdS/CFT correspondence. In this we claim that one can use Spin Matrix theory as a connecting link to non-perturbative effects in string theory on $\ads_5\times S^5$. An immediate objection to this is that we are taking the $\lambda\rightarrow 0$ limit in approaching a zero-temperature critical point, which seems at odds with connecting to the string theory side at large $\lambda$. However, as explained in Section \ref{sec:g_infty}, it is already clear that one can match the $g \rightarrow \infty$ limit of Spin Matrix theory to the supersymmetric sector of $\ads_5\times S^5$. In itself this can already be a very useful result since in this paper we have found classical regimes for $SU(2)$ Spin Matrix theory which makes it simple to take the $g\rightarrow \infty$ limit, and one could very well imagine finding similar classical regimes for other Spin Matrix theories. Furthermore, as we will mention below, there are important unresolved issues in the supersymmetric sector of the AdS/CFT correspondence.

We believe that one can employ Spin Matrix theory as a connecting link to string theory beyond the supersymmetric sector. Again, the challenge here is that we seemingly are deep into the quantum regime on the string theory side by taking $\lambda \rightarrow 0$. However, we propose that large $\lambda$ is not always a necessity on the string theory side of the AdS/CFT correspondence. Indeed, what one more precisely should consider is whether one is in a semi-classical regime on the string theory side, or not. The Spin Matrix theory limits can be used to identify regimes on the string theory which are semi-classical, even when the 't Hooft coupling is small. By considering the string theory side in such regimes, one can take $\lambda \rightarrow 0$ and still have a large action. One can then compare this action to the corresponding regime on the gauge theory side, which then is in terms of the Spin Matrix theory. Our assertion is that by having large $g$ in Spin Matrix theory one should be able to completely, or at least closely, match to the mentioned semi-classical regime on the string theory side. As already noted the leading $g \rightarrow \infty$ behavior is protected by supersymmetry, and hence this should suppress zero-mode quantum fluctuations. Moreover, the modes that decouple in the Spin Matrix theory limit become infinitely heavy. Indeed, we have shown in \cite{Harmark:2008gm} that this reasoning works well for tree-level string theory ($i.e.$ the planar limit). 

On a more general note we would like to emphasize that it seems to us that one can employ Spin Matrix theory to identify (semi-)classical regions on both sides of the correspondence. Thus, even if one considers examples where the above reasoning does not hold, we believe that having dual classical regions to match between is a very good starting point for building a more detailed match.

\subsubsection*{Open problems}

One of the most interesting future directions is to study Spin Matrix theories with non-compact spin group (see Table \ref{tab:SMT}). For these Spin Matrix theories one would get a quantum mechanical theory that for large temperatures effectively has a number of continuous directions, $e.g.$ for the $SU(1,2|3)$ Spin Matrix theory one would have two continous directions. One should be able to find an analog of the classical matrix model of Section \ref{sec:clas_any} that would have dependence on spatial directions as well. $SU(1,2|3)$ Spin Matrix theory is particularly interesting since it should contain the $1/16$ BPS supersymmetric states that are dual to black holes on the string side of the AdS/CFT correspondence \cite{Gutowski:2004ez} %,Gutowski:2004yv}
 (since they satisfy $M = S_1+S_2+R_1+R_2+R_3$ where $M$ is the mass of the black hole). In this case one should be able to relate the $g\rightarrow \infty$ and large $T$ regime of $SU(1,2|3)$ Spin Matrix theory to the black hole thermodynamics (and possibly also for $g$ large). Understanding the $1/16$ BPS states that underlies these supersymmetric black holes from the point of view of $\CN=4$ SYM is an outstanding problem in the literature that has proven to be quite difficult to solve \cite{Grant:2008sk,Chang:2013fba}.

Regarding the high-temperature phase of $SU(2)$ Spin Matrix theory one of our key results is that free $SU(q)$ Spin Matrix theory at high temperature behaves like $(q-1)N^2+1$ uncoupled one-dimensional harmonic oscillators. As remarked in Section \ref{sec:suq_partfcts} it is unclear what the underlying description in terms of the states of the Hilbert space is for these emerging harmonic oscillator degrees of freedom. Indeed, even in the classical description using coherent states of Section \ref{sec:classical_phase_zero} it is unclear. This would be interesting to study further in order to clarify the emergence of these harmonic oscillators. We emphasize that one of the interesting features is that these emerging degrees of freedom are uncoupled.

Another key result for the high-temperature phase of $SU(2)$ Spin Matrix theory is the classical matrix model description for any coupling $g$, as well as the result that for large coupling $g \rightarrow \infty$ one obtains a phase described by $2N$ uncoupled one-dimensional harmonic oscillators. As remarked in Section \ref{sec:clas_any} the classical matrix model is the high-temperature equivalent to the Heisenberg spin chain of a connecting link between weak and strong coupling, and it would be highly interesting to study if and how such a matrix model can be obtained on the string theory side of the AdS/CFT correspondence (see Section \ref{sec:clas_any} for remarks on this). 

One can easily translate the results of Section \ref{sec:hightemp} to $SU(2)$ Spin Matrix theory based on the adjoint representation of $SU(N)$ (rather than $U(N)$). Starting from the classical matrix model of Section \ref{sec:clas_any} one notices that the traces of the four matrices $X_s$ and $P_s$ decouple from the other matrix components. The $SU(N)$ case corresponds to setting these traces to zero, which means one should remove two oscillators at high temperature. Hence for $g=0$ we obtain instead $N^2-1$ harmonic oscillators while for $g\rightarrow\infty$ we have $2N-2$ harmonic oscillators.

There are several other interesting problems to investigate for $SU(2)$ Spin Matrix theory. One can turn on the chemical potential $\mu$ conjugate to $S_z$ which measures the total spin. For the low temperature phases this was investigated in \cite{Harmark:2006ie}. It could be interesting to examine the high temperature phases for non-zero chemical potential as it would reveal more information on the nature of the emergent $N^2 +1$ harmonic oscillators for $g=0$. Instead for non-zero coupling one would presumably find a straightforward generalization of the classical matrix model of Section \ref{sec:clas_any}.

Furthermore, one could investigate free $SU(q)$ Spin Matrix theory by taking a large $q$ limit. In this limit the leading large $q$ behavior could be classical which could help shed some light on the high temperature phase of uncoupled harmonic oscillators. 

It could also be interesting to explore the connection to the work of \cite{Carlson:2011hy} where evidence for integrability symmetry in excitations of $\CN=4$ operators dual to Giant Gravitons is found. For the one-loop dilatation operator this is a result which is part of $SU(2)$ Spin Matrix theory and hence one could explore this further within this framework.

Finally, one can investigate the partial deconfinement transition that separates the low- and high-temperature phases, $i.e.$ the phase transition that occurs at $T_c$, where we go from confining $\log Z \sim \CO(1)$ behavior to deconfining $\log Z \sim \CO(N^2)$ behavior. While for $g=0$ one finds that the phase transition occurs at the Hagedorn temperature $T_H = 1/\log 2$ it is likely that $T_c$ lies lower than the Hagedorn temperature above $g>0$ in which case this phase transition would be a first order transition. However, as pointed out in \cite{Aharony:2003sx}, one can also first encounter the Hagedorn phase transition as a second order phase transition and subsequently a continuous phase transition at a higher temperature into the high-temperature phase.

%%%%%%%%%%%%%%%%%%%%%%%%%%%%%%
\section*{Acknowledgments}

We thank Robert de Mello Koch, Jelle Hartong, Yang-Hui He, Niels Obers and Sanjaye Ramgoolam for useful discussions and comments. TH acknowledges support from the ERC-advance grant ``Exploring the Quantum Universe" as well as from the Marie-Curie-CIG grant ``Quantum Mechanical Nature of Black Holes" both from the European Union. MO thanks the Niels Bohr Institute for hospitality and acknowledges support from Museo Storico della Fisica e Centro Studi e Ricerche Enrico Fermi while part of this work was under preparation.

\begin{appendix}

%%%%%%%%%%%%%%%%%%%%%%%%%%%%%%%%%%%%%%%%%%%%%%%%%%%%%%%%%%
\section{Partition functions of free $SU(q)$ Spin Matrix theory}
\label{app:partfcts}

Using Eq.~\eqref{freepartfct2} we have computed the partition function of free $SU(q)$ Spin Matrix theory for general $q$ and for $2 \leq N \leq 5$ up to order $x^{40}$. Employing the assumption that the partition functions are of the form $P(x)/Q(x)$ we have subsequently resummed the series for specific values of $(q,N)$. With $N=2$ we have resummed the series for $2 \leq q \leq 5$, with $N=3$ for $2 \leq q \leq 5$, with $N=4$ for $q=2$ and with $N=5$ for $q=2$. The resummed expressions for the partition functions are%
\footnote{$Z_{2,2}(\beta)|_{g=0}$ and $Z_{2,3}(\beta)|_{g=0}$ have previously been computed in \cite{Kimura:2009ur}.} 
\begin{equation}
\label{Zq2}
Z_{q,2}(\beta)|_{g=0} = \frac{P_{2,2q-4}(x)}{(1-x)^{2q-2} (1-x^2)^{2q-1}} \ \ \mbox{for} \ \ 2 \leq q \leq 5
\end{equation}
\begin{equation}
Z_{q,3}(\beta)|_{g=0} = \frac{P_{3,10q-16}(x)}{(1-x)^{2q-2} (1-x^2)^{4q-4}(1-x^3)^{3q-2}}  \ \ \mbox{for} \ \ 2 \leq q \leq 5
\end{equation}
\begin{equation}
Z_{2,4}(\beta)|_{g=0} = \frac{P_{4,14}(x)}{(1-x)^3(1-x^2)^4(1-x^3)^5(1-x^4)^5}
\end{equation}
\begin{equation}
\label{Z25}
Z_{2,5}(\beta)|_{g=0} = \frac{P_{5,39}(x)}{(1-x^2)^6 (1-x^3)^8 (1-x^4)^6 (1-x^5)^6} 
\end{equation}
with the polynomials 
\begin{equation}
\begin{array}{c}
P_{2,0} = 1 \spa P_{2,2} = 1-x+x^2 \spa P_{2,4} = 1-2x+4x^2-2x^3+x^4 \\[2mm] P_{2,6} = 1-3x+9x^2-9x^3+9x^4-3x^5+x^6
\end{array}
\end{equation}
for $N=2$ and
\begin{equation}
\begin{array}{c}
P_{3,4} = 1-x^2+x^4 
\\[2mm]
\begin{array}{rl}
P_{3,14} = &1-x-2x^2+6x^3+6x^4-9x^5+x^6+17x^7+x^8-9x^9 \\ &+6x^{10}+6x^{11}-2x^{12}-x^{13}+x^{14} 
\end{array}
\\[5mm]
\begin{array}{rl}
P_{3,24} = & 1-2x-x^2+18x^3+6x^4-30x^5+75x^6+150x^7-30x^8+30x^9  \\ &+410x^{10}+238x^{11}-76x^{12}+238x^{13}+401x^{14}+30x^{15}-30x^{16}  \\ &+150x^{17}+75x^{18}-30x^{19}+6x^{20}+18x^{21}-x^{22}-2x^{23}+x^{24}
\end{array}
\\[8mm]
\begin{array}{rl}
P_{3,34} = & 1-3x+2x^2+34x^3-4x^4-18x^4+421x^6+624x^7+251x^8 \\  & +2107x^9+5377x^{10}+4766x^{11} +6384x^{12}+16031x^{13}+19327x^{14} \\ &+14592x^{15}  +21381x^{16} +29839x^{17}+21381x^{18}+14592x^{19}+19327x^{20} \\ & +16031x^{21} +6384x^{22}+4766x^{23}+5377x^{24}+2107x^{25}+251x^{26} \\ & +624x^{27}+421x^{28} -18x^{29}-4x^{30}+34x^{31}+2x^{32}-3x^{33}+x^{34}   
\end{array}
\end{array}
\end{equation}
for $N=3$ and
\begin{equation}
\begin{array}{c}
P_{4,14} =  1-x-x^2+2x^4+2x^5-4x^7+2x^9 +2x^{10}-x^{12}-x^{13}+x^{14}
\\[2mm]
\begin{array}{rl}
P_{5,39} = & 1+2x-6x^3-9x^4+2x^5+25x^6+38x^7+17x^8-34x^9-68x^{10} \\ & -34x^{11}+73x^{12}+176x^{13}+171x^{14}+34x^{15}-127x^{16}-156x^{17}-2x^{18} \\ &+218x^{19}+322x^{20}+218x^{21}-2x^{22}-156x^{23}-127x^{24}+34x^{25} \\ & +171x^{26}+176x^{27}   +73x^{28} -34x^{29} -68x^{30} -34x^{31} +17x^{32} +38x^{33} \\ & +26x^{34} +2x^{35} -9x^{36} -6x^{37} +2x^{38} +x^{39}
\end{array}
\end{array}
\end{equation}
for $N=4$ and $N=5$. Notice that all of the polynomials are palindromic.

For $x \rightarrow 1$ all the resummed partition functions Eqs.~\eqref{Zq2}-\eqref{Z25} have a limit of the form
\begin{equation}
Z_{q,N} (\beta)|_{g=0} \simeq \frac{a_{q,N}}{(1-x)^{(q-1)N^2+1}} \ \ \mbox{for} \ \ T \rightarrow \infty
\end{equation}
where the coefficient $a_{q,N}$ is given in Table \ref{tab:aqN}.

\begin{table}[ht]
\begin{center}
\begin{tabular}{|c||c|c|c|c|}
\hline $a_{q,N}$ & $q=2$ & $q=3$ & $q=4$ & $q=5$ \\ 
\hline \hline $N=2$ & $\frac{1}{2^3}$ & $\frac{1}{2^{5}}$ & $\frac{1}{2^6}$ & $\frac{5}{2^9}$ \\
\hline $N=3$ & $\frac{1}{2^4 3^4}$ & $\frac{7}{2^8 3^6}$ & $\frac{409}{2^{10}3^{10}}$ & $\frac{5 \cdot 14159}{2^{16}3^{12}}$ \\
\hline $N=4$ & $\frac{1}{2^{13}3^5}$ &  &  &  \\
\hline $N=5$ & $\frac{193}{2^{18} 3^8 5^5}$ &  &  &  \\
\hline
\end{tabular}
\caption{Table of $a_{q,N}$ coefficients. \label{tab:aqN}}
\end{center}
\end{table}

The Plethystic logarithm of a partition function $Z(x)$ is defined as \cite{Benvenuti:2006qr}
\begin{equation}
(PE^{-1} ( Z ))(x) = \sum_{k=1}^\infty \frac{\mu(k)}{k} \log Z (x^k ) 
\end{equation}
where $\mu(k)$ is the Mobius function which is $0$ for repeated primes, $1$ for $k=1$ and $(-1)^n$ when $k$ is a product of $n$ distinct primes. Considering the Plethystic logarithms of the partition functions Eqs.~\eqref{Zq2}-\eqref{Z25} we find that the following three cases are finite%
\footnote{$PE^{-1}(Z_{2,2})$ and $PE^{-1}(Z_{2,3})$ have previously been computed in \cite{Kimura:2009ur}.} 
\begin{equation}
\label{plethy_res}
\begin{array}{c}
PE^{-1}(Z_{2,2}) = 2x+3x^2 \spa PE^{-1} (Z_{3,2}) = 3x+6x^2+x^3-x^6 
\\[2mm] PE^{-1} (Z_{2,3}) = 2x+3x^2+4x^3+x^4+x^6-x^{12}
\end{array}
\end{equation}
For all the other cases, so for $N=2$ with $q=4,5$, $N=3$ with $q=3,4,5$ and $(q,N)=(2,4),(2,5)$, one can see that the Plethystic logarithm gives an infinite series in $x$. This follows from the fact that one cannot put the partition functions in the so-called Euler form $\prod_{k=1}^\infty (1-x^k)^{-b_k}$ with a finite number of non-zero integers $b_k$ \cite{Benvenuti:2006qr}.

For our purposes the interpretation of the Plethystic logarithm is that given a partition function $Z(x)$ over multi-trace states the Plethystic logarithm returns the corresponding single-trace partition function $Z_{\rm ST}(x)$ from which one can generate $Z(x)$, hence \cite{Benvenuti:2006qr}
\begin{equation}
\log Z(x) = \sum_{k=1}^\infty \frac{1}{k} Z_{\rm ST} (x^k) \spa Z_{\rm ST}(x) = (PE^{-1}(Z))(x)
\end{equation}
This relation requires that the Hamiltonian does not affect the multi-trace structure when acting on a state. This is indeed the case for the partition functions \eqref{Zq2}-\eqref{Z25} since we are considering $g=0$.

The result $PE^{-1}(Z_{2,2}) = 2x+3x^2$ means that one can generate the full Hilbert space by combining two single-traces of length one ($\tr(a^\dagger_\uparrow)$ and $\tr(a^\dagger_\downarrow)$) and three single-traces of length two ($\tr(a^\dagger_\uparrow a^\dagger_\uparrow)$, $\tr(a^\dagger_\uparrow a^\dagger_\downarrow)$ and $\tr(a^\dagger_\downarrow a^\dagger_\downarrow)$). For $PE^{-1}(Z_{3,2})$ in \eqref{plethy_res} it means one has single-traces up to length 3 but in addition an algebraic relation of length 6 between these single-traces.

%%%%%%%%%%%%%%%%%%%%%%%%%%%%%%%%%%%%%%%%%%%%%%%%%%%%%%%%%%
\section{Free $SU(q)$ Spin Matrix theory using eigenvalue description}
\label{app:eigenval}

In Section \ref{sec:suq_partfcts} we conjecture that the partition function \eqref{freepartfct3} for free $SU(q)$ Spin Matrix theory in general has the high temperature behavior \eqref{largeTpre} based on computations of the partition function for low values of $N$. In this appendix we check this result in the large $N$ regime by applying the method of eigenvalues for the unitary matrix $U$ in \eqref{freepartfct1} with \eqref{smallz} (see for instance Ref.~\cite{Aharony:2003sx}). In this approach we write $e^{i\theta_j}$ as the $N$ eigenvalues of $U$. Then the partition function \eqref{freepartfct3} can be written for large $N$ as 
\begin{equation}
Z_{q,N}(\beta) |_{g=0} = \int \prod_i [ d\theta_i] e^{-I} \spa I = 2\pi N^2 \sum_{n=1} \frac{1}{n} (1-qx^n) |\rho_n|^2 
\end{equation}
where $I[\rho(\theta)]$ is the effective action for a continuous distribution $\rho (\theta)$ of eigenvalues with $\int_{-\pi}^\pi d\theta \rho(\theta) =1$ and $\rho_n = \int_{-\pi}^\pi d\theta \rho(\theta) \cos (n\theta)$ where we assume without loss of generality that the distribution is symmetric around $\theta=0$. For temperatures above the Hagedorn temperature $T_H = 1/ \log q$ one has $\log Z_{q,N}(\beta)|_{g=0} = - I_{\rm min}(\beta)$ where $I_{\rm min}(\beta)$ is the minimum of the effective action. 

\begin{figure}[h!]
\centerline{\includegraphics[scale=0.7]{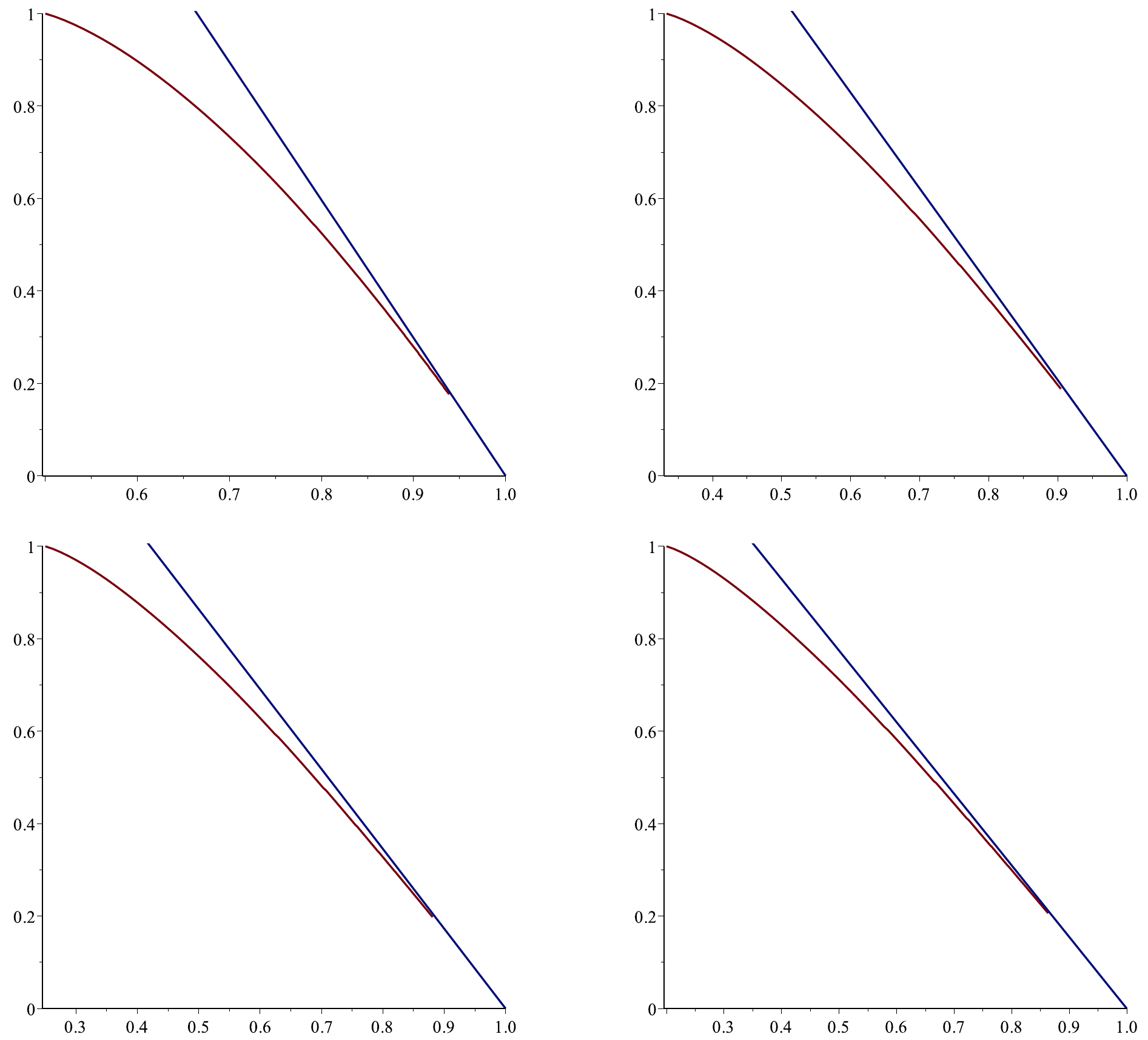}}
\caption{\small Above $L_q(x)$ is plotted as function of $x$ for $q=2,3,4,5$. These plots are found numerically starting from $x=1/q$ and ending at $x=(1/q)^{1/11}$. The linear curve $(b_q)^{\frac{1}{q-1}} ( 1-x )$ has been plotted as estimate for the behavior of $L_q(x)$ for $x \rightarrow 1$.}
\label{fig:Lq}
\begin{picture}(0,0)(0,0)
\put(208,253){$x$}
\put(415,253){$x$}
\put(208,75){$x$}
\put(415,75){$x$}
\put(20,328){$L_2$}
\put(227,328){$L_3$}
\put(20,150){$L_4$}
\put(227,150){$L_5$}
\put(160,378){$q=2$}
\put(367,378){$q=3$}
\put(160,200){$q=4$}
\put(367,200){$q=5$}
\end{picture}
\end{figure}
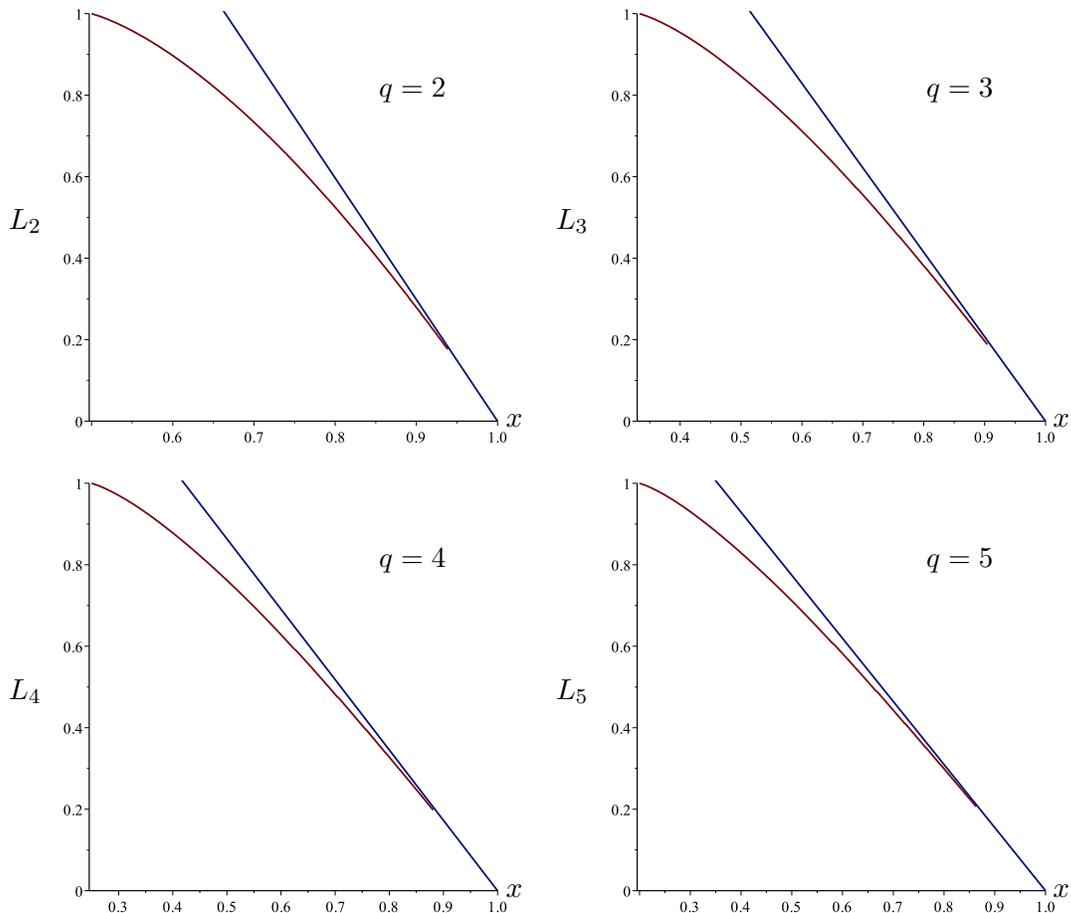

The method presented in Ref.~\cite{Aharony:2003sx} to find $I_{\rm min}(\beta)$ to a good approximation is to include the first $k$ modes $\rho_1,...,\rho_k$, and setting the rest to zero $\rho_n=0$ for $n > k$. However, since we want to approach $x \rightarrow 1$ we have to choose $k$ with care. Namely, each time $x$ passes a point $(1/q)^{1/n}$ for some $n$ a new mode $\rho_n$ becomes massless and hence cannot be ignored for larger $x$. To accomodate this we increase $k$ with one each time we pass such a point. We find that one obtains reliable results starting with $k=4$ at $x=1/q$. We then compute $L_q(x)$ up to the value $x=(1/q)^{1/11}$ with $k=13$ in the final interval. According to the conjecture \eqref{largeTpre} one should have for large $N$
\begin{equation}
\label{Lqdef}
L_{q}(x) \equiv \lim_{N \rightarrow \infty} \exp \left( - \frac{\log Z_{q,N}(\beta) |_{g=0}}{(q-1)N^2} \right) \simeq (b_q)^{\frac{1}{q-1}} ( 1-x ) \ \ \mbox{for} \ \ x \rightarrow 1
\end{equation}
where $b_q$ is a constant which is subleading for $x\rightarrow 1$. Thus, if the quantity $L_{q}(x)$ goes to zero approximately linearly for $x\rightarrow 1$ then it is in line with the conjecture \eqref{largeTpre}.
In Fig.~\ref{fig:Lq} we have plotted $L_{q}(x)$ for $q=2,3,4,5$. We see that one indeed gets a linear behavior  of the form $(b_q)^{\frac{1}{q-1}} ( 1-x )$ as $x\rightarrow 1$, thus giving evidence to the conjecture \eqref{largeTpre}.

The constant $b_q$ in \eqref{Lqdef} is related to the constant $a_{q,N}$ in \eqref{largeTpre} as follows
\begin{equation}
b_q = \lim_{N\rightarrow \infty} \exp \left( - \frac{\log a_{q,N}}{N^2} \right) 
\end{equation}
We find graphically in the diagrams of Fig.~\ref{fig:Lq} that $b_q \simeq 8.9$ with uncertainty $\pm 0.05$.%
\footnote{If one takes the values for $a_{2,N}$ of Table \ref{tab:aqN} for $N=2,3,4,5$ one finds $(a_{2,2})^{-1/2^2} \simeq  1.7$, $(a_{2,3})^{-1/3^2}\simeq 2.2$, $(a_{2,4})^{-1/4^2}\simeq 2.5$ and $(a_{2,5})^{-1/5^2}\simeq 2.6$. This behavior is consistent with what we find for $b_2$.} 
 Note that we find the same value for $b_q$ for all $q=2,3,4,5$ within the uncertainty. It could be interesting to pursue this further in order to understand if $b_q$ is the same for all $q$.

\end{appendix}

%%%%%%%%%%%%%%%%%%%%%%%%%%%%%%%%%%%%%%%%%%%%%%%%%%%%%%%%%%%%%

\small

%\bibliographystyle{newutphys}
%\bibliography{mybib}

\begin{thebibliography}{10}

\bibitem{Maldacena:1997re}
J.~M. Maldacena, ``{The large N limit of superconformal field theories and
  supergravity},'' {\em Adv. Theor. Math. Phys.} {\bf 2} (1998)  231--252,
\href{http://arxiv.org/abs/hep-th/9711200}{{\tt arXiv:hep-th/9711200}}.
%%CITATION = HEP-TH/9711200;%%.
%
%\bibitem{Gubser:1998bc}
S.~S. Gubser, I.~R. Klebanov, and A.~M. Polyakov, ``{Gauge theory correlators
  from non-critical string theory},''
  \href{http://dx.doi.org/10.1016/S0370-2693(98)00377-3}{{\em Phys. Lett.} {\bf
  B428} (1998)  105--114},
\href{http://arxiv.org/abs/hep-th/9802109}{{\tt arXiv:hep-th/9802109}}.
%%CITATION = HEP-TH/9802109;%%.
%
%\bibitem{Witten:1998qj}
E.~Witten, ``{Anti-de Sitter space and holography},'' {\em Adv. Theor. Math.
  Phys.} {\bf 2} (1998)  253--291,
\href{http://arxiv.org/abs/hep-th/9802150}{{\tt arXiv:hep-th/9802150}}.
%%CITATION = HEP-TH/9802150;%%.

\bibitem{Pestun:2007rz}
V.~Pestun, ``{Localization of gauge theory on a four-sphere and supersymmetric
  Wilson loops},'' \href{http://dx.doi.org/10.1007/s00220-012-1485-0}{{\em
  Commun.Math.Phys.} {\bf 313} (2012)  71--129},
\href{http://arxiv.org/abs/0712.2824}{{\tt arXiv:0712.2824 [hep-th]}}.
%%CITATION = ARXIV:0712.2824;%%.

\bibitem{Minahan:2002ve}
J.~A. Minahan and K.~Zarembo, ``The {Bethe-ansatz} for {$\CN = 4$} super
  {Yang-Mills},'' {\em JHEP} {\bf 03} (2003)  013,
\href{http://arxiv.org/abs/hep-th/0212208}{{\tt hep-th/0212208}}.
%%CITATION = HEP-TH 0212208;%%.

\bibitem{Beisert:2003tq}
N.~Beisert, C.~Kristjansen, and M.~Staudacher, ``The dilatation operator of
  {$\CN = 4$} super {Yang-Mills} theory,'' {\em Nucl. Phys.} {\bf B664} (2003)
  131--184,
\href{http://arxiv.org/abs/hep-th/0303060}{{\tt hep-th/0303060}}.
%%CITATION = HEP-TH 0303060;%%.

\bibitem{Beisert:2006ez}
N.~Beisert, B.~Eden, and M.~Staudacher, ``{Transcendentality and crossing},''
  {\em J. Stat. Mech.} {\bf 0701} (2007)  P021,
\href{http://arxiv.org/abs/hep-th/0610251}{{\tt hep-th/0610251}}.
%%CITATION = HEP-TH/0610251;%%.

\bibitem{Carlson:2011hy}
W.~Carlson, R.~d.~M. Koch, and H.~Lin, ``{Nonplanar Integrability},''
  \href{http://dx.doi.org/10.1007/JHEP03(2011)105}{{\em JHEP} {\bf 1103} (2011)
   105},
\href{http://arxiv.org/abs/1101.5404}{{\tt arXiv:1101.5404 [hep-th]}}.
%%CITATION = ARXIV:1101.5404;%%.
%
%\bibitem{Koch:2011jk}
R.~d.~M. Koch, B.~A.~E. Mohammed, and S.~Smith, ``{Nonplanar Integrability:
  Beyond the SU(2) Sector},''
  \href{http://dx.doi.org/10.1142/S0217751X11054590}{{\em Int.J.Mod.Phys.} {\bf
  A26} (2011)  4553--4583},
\href{http://arxiv.org/abs/1106.2483}{{\tt arXiv:1106.2483 [hep-th]}}.
%%CITATION = ARXIV:1106.2483;%%.

\bibitem{Kristjansen:2010kg}
C.~Kristjansen, ``{Review of AdS/CFT Integrability, Chapter IV.1: Aspects of
  Non-Planarity},'' \href{http://dx.doi.org/10.1007/s11005-011-0514-9}{{\em
  Lett.Math.Phys.} {\bf 99} (2012)  349--374},
\href{http://arxiv.org/abs/1012.3997}{{\tt arXiv:1012.3997 [hep-th]}}.
%%CITATION = ARXIV:1012.3997;%%.

\bibitem{Harmark:2006di}
T.~Harmark and M.~Orselli, ``Quantum mechanical sectors in thermal {$\CN = 4$}
  super {Yang-Mills} on {$\R \times S^3$},'' {\em Nucl. Phys.} {\bf B757}
  (2006)  117--145,
\href{http://arxiv.org/abs/hep-th/0605234}{{\tt hep-th/0605234}}.
%%CITATION = HEP-TH 0605234;%%.

\bibitem{Harmark:2006ta}
T.~Harmark and M.~Orselli, ``Matching the {Hagedorn} temperature in
  {AdS/CFT},'' {\em Phys. Rev.} {\bf D74} (2006)  126009,
\href{http://arxiv.org/abs/hep-th/0608115}{{\tt hep-th/0608115}}.
%%CITATION = HEP-TH 0608115;%%.

\bibitem{Harmark:2006ie}
T.~Harmark, K.~R. Kristjansson, and M.~Orselli, ``Magnetic {Heisenberg-chain} /
  pp-wave correspondence,'' {\em JHEP} {\bf 02} (2007)  085,
\href{http://arxiv.org/abs/hep-th/0611242}{{\tt hep-th/0611242}}.
%%CITATION = HEP-TH/0611242;%%.

\bibitem{Harmark:2007px}
T.~Harmark, K.~R. Kristjansson, and M.~Orselli, ``Decoupling limits of
  {$\CN=4$} super {Yang-Mills} on {$\R \times S^3$},''
  \href{http://dx.doi.org/10.1088/1126-6708/2007/09/115}{{\em JHEP} {\bf 09}
  (2007)  115},
\href{http://arxiv.org/abs/0707.1621}{{\tt arXiv:0707.1621 [hep-th]}}.
%%CITATION = 0707.1621;%%.

\bibitem{Harmark:2008gm}
T.~Harmark, K.~R. Kristjansson, and M.~Orselli, ``{Matching gauge theory and
  string theory in a decoupling limit of AdS/CFT},''
  \href{http://dx.doi.org/10.1088/1126-6708/2009/02/027}{{\em JHEP} {\bf 0902}
  (2009)  027},
\href{http://arxiv.org/abs/0806.3370}{{\tt arXiv:0806.3370 [hep-th]}}.
%%CITATION = ARXIV:0806.3370;%%.

\bibitem{Kruczenski:2003gt}
M.~Kruczenski, ``Spin chains and string theory,'' {\em Phys. Rev. Lett.} {\bf
  93} (2004)  161602,
\href{http://arxiv.org/abs/hep-th/0311203}{{\tt hep-th/0311203}}.
%%CITATION = HEP-TH 0311203;%%.

\bibitem{Hanada:2013rga}
M.~Hanada, Y.~Hyakutake, G.~Ishiki, and J.~Nishimura, ``{Holographic
  description of quantum black hole on a computer},''
\href{http://arxiv.org/abs/1311.5607}{{\tt arXiv:1311.5607 [hep-th]}}.
%%CITATION = ARXIV:1311.5607;%%.

\bibitem{Balasubramanian:2004nb}
V.~Balasubramanian, D.~Berenstein, B.~Feng, and M.-x. Huang, ``{D-branes in
  Yang-Mills theory and emergent gauge symmetry},''
  \href{http://dx.doi.org/10.1088/1126-6708/2005/03/006}{{\em JHEP} {\bf 0503}
  (2005)  006},
\href{http://arxiv.org/abs/hep-th/0411205}{{\tt arXiv:hep-th/0411205
  [hep-th]}}.
%%CITATION = HEP-TH/0411205;%%.

\bibitem{Koch:2007uu}
R.~de~Mello~Koch, J.~Smolic, and M.~Smolic, ``{Giant Gravitons - with Strings
  Attached (I)},'' \href{http://dx.doi.org/10.1088/1126-6708/2007/06/074}{{\em
  JHEP} {\bf 0706} (2007)  074},
\href{http://arxiv.org/abs/hep-th/0701066}{{\tt arXiv:hep-th/0701066
  [hep-th]}}.
%%CITATION = HEP-TH/0701066;%%.

\bibitem{Sundborg:1999ue}
B.~Sundborg, ``The {Hagedorn} transition, deconfinement and {$\CN = 4$ SYM}
  theory,'' {\em Nucl. Phys.} {\bf B573} (2000)  349--363,
\href{http://arxiv.org/abs/hep-th/9908001}{{\tt hep-th/9908001}}.
%%CITATION = HEP-TH 9908001;%%.

\bibitem{Aharony:2003sx}
O.~Aharony, J.~Marsano, S.~Minwalla, K.~Papadodimas, and M.~Van~Raamsdonk,
  ``{The Hagedorn / deconfinement phase transition in weakly coupled large N
  gauge theories},'' {\em Adv. Theor. Math. Phys.} {\bf 8} (2004)  603--696,
\href{http://arxiv.org/abs/hep-th/0310285}{{\tt arXiv:hep-th/0310285}}.
%%CITATION = HEP-TH/0310285;%%.

\bibitem{Dolan:2002zh}
F.~A. Dolan and H.~Osborn, ``On short and semi-short representations for four
  dimensional superconformal symmetry,'' {\em Ann. Phys.} {\bf 307} (2003)
  41--89,
\href{http://arxiv.org/abs/hep-th/0209056}{{\tt hep-th/0209056}}.
%%CITATION = HEP-TH 0209056;%%.

\bibitem{Bianchi:2003wx}
M.~Bianchi, J.~F. Morales, and H.~Samtleben, ``On stringy {AdS}{$_5 \times
  S^5$} and higher spin holography,'' {\em JHEP} {\bf 07} (2003)  062,
\href{http://arxiv.org/abs/hep-th/0305052}{{\tt hep-th/0305052}}.
%%CITATION = HEP-TH 0305052;%%.

\bibitem{Beisert:2003jj}
N.~Beisert, ``The complete one-loop dilatation operator of {$\CN = 4$} super
  {Yang-Mills} theory,'' {\em Nucl. Phys.} {\bf B676} (2004)  3--42,
\href{http://arxiv.org/abs/hep-th/0307015}{{\tt hep-th/0307015}}.
%%CITATION = HEP-TH 0307015;%%.

\bibitem{Beisert:2005tm}
N.~Beisert, ``{The $SU(2|2)$ dynamic S-matrix},''
  \href{http://dx.doi.org/10.4310/ATMP.2008.v12.n5.a1}{{\em
  Adv.Theor.Math.Phys.} {\bf 12} (2008)  945--979},
\href{http://arxiv.org/abs/hep-th/0511082}{{\tt arXiv:hep-th/0511082
  [hep-th]}}.
%%CITATION = HEP-TH/0511082;%%.

\bibitem{Berenstein:2002jq}
D.~Berenstein, J.~M. Maldacena, and H.~Nastase, ``Strings in flat space and pp
  waves from {$\CN = 4$} super {Yang Mills},'' {\em JHEP} {\bf 04} (2002)  013,
\href{http://arxiv.org/abs/hep-th/0202021}{{\tt hep-th/0202021}}.
%%CITATION = HEP-TH 0202021;%%.

\bibitem{Beisert:2004ry}
N.~Beisert, ``The dilatation operator of {$\CN = 4$} super {Yang-Mills} theory
  and integrability,'' {\em Phys. Rept.} {\bf 405} (2005)  1--202,
\href{http://arxiv.org/abs/hep-th/0407277}{{\tt hep-th/0407277}}.
%%CITATION = HEP-TH 0407277;%%.

\bibitem{Kinney:2005ej}
J.~Kinney, J.~M. Maldacena, S.~Minwalla, and S.~Raju, ``{An Index for 4
  dimensional super conformal theories},''
  \href{http://dx.doi.org/10.1007/s00220-007-0258-7}{{\em Commun.Math.Phys.}
  {\bf 275} (2007)  209--254},
\href{http://arxiv.org/abs/hep-th/0510251}{{\tt arXiv:hep-th/0510251
  [hep-th]}}.
%%CITATION = HEP-TH/0510251;%%.

\bibitem{Berkooz:2006wc}
M.~Berkooz, D.~Reichmann, and J.~Simon, ``{A Fermi Surface Model for Large
  Supersymmetric $\ads_5$ Black Holes},''
  \href{http://dx.doi.org/10.1088/1126-6708/2007/01/048}{{\em JHEP} {\bf 0701}
  (2007)  048},
\href{http://arxiv.org/abs/hep-th/0604023}{{\tt arXiv:hep-th/0604023
  [hep-th]}}.
%%CITATION = HEP-TH/0604023;%%.

\bibitem{Grant:2008sk}
L.~Grant, P.~A. Grassi, S.~Kim, and S.~Minwalla, ``{Comments on 1/16 BPS
  Quantum States and Classical Configurations},''
  \href{http://dx.doi.org/10.1088/1126-6708/2008/05/049}{{\em JHEP} {\bf 0805}
  (2008)  049},
\href{http://arxiv.org/abs/0803.4183}{{\tt arXiv:0803.4183 [hep-th]}}.
%%CITATION = ARXIV:0803.4183;%%.

\bibitem{Biswas:2006tj}
I.~Biswas, D.~Gaiotto, S.~Lahiri, and S.~Minwalla, ``{Supersymmetric states of
  N=4 Yang-Mills from giant gravitons},''
  \href{http://dx.doi.org/10.1088/1126-6708/2007/12/006}{{\em JHEP} {\bf 0712}
  (2007)  006},
\href{http://arxiv.org/abs/hep-th/0606087}{{\tt arXiv:hep-th/0606087
  [hep-th]}}.
%%CITATION = HEP-TH/0606087;%%.

\bibitem{Fradkin}
{See for example chapter 5 in: E. Fradkin}, ``{Field theories of condensed
  matter systems},'' {\em Addison-Wesley Publishing Company, Redwood City, CA}
  (1991)  .

\bibitem{Bertolini:2002nr}
M.~Bertolini, J.~de~Boer, T.~Harmark, E.~Imeroni, and N.~A. Obers, ``Gauge
  theory description of compactified pp-waves,'' {\em JHEP} {\bf 01} (2003)
  016,
\href{http://arxiv.org/abs/hep-th/0209201}{{\tt hep-th/0209201}}.
%%CITATION = HEP-TH 0209201;%%.

\bibitem{Bellucci:2004ru}
S.~Bellucci, P.~Casteill, J.~Morales, and C.~Sochichiu, ``{Spin bit models from
  nonplanar N = 4 SYM},''
  \href{http://dx.doi.org/10.1016/j.nuclphysb.2004.07.025}{{\em Nucl.Phys.}
  {\bf B699} (2004)  151--173},
\href{http://arxiv.org/abs/hep-th/0404066}{{\tt arXiv:hep-th/0404066
  [hep-th]}}.
%%CITATION = HEP-TH/0404066;%%.

\bibitem{Peeters:2004pt}
K.~Peeters, J.~Plefka, and M.~Zamaklar, ``{Splitting spinning strings in
  AdS/CFT},'' \href{http://dx.doi.org/10.1088/1126-6708/2004/11/054}{{\em JHEP}
  {\bf 0411} (2004)  054},
\href{http://arxiv.org/abs/hep-th/0410275}{{\tt arXiv:hep-th/0410275
  [hep-th]}}.
%%CITATION = HEP-TH/0410275;%%.

\bibitem{Casteill:2007td}
P.~Y. Casteill, R.~A. Janik, A.~Jarosz, and C.~Kristjansen, ``{Quasilocality of
  joining/splitting strings from coherent states},''
  \href{http://dx.doi.org/10.1088/1126-6708/2007/12/069}{{\em JHEP} {\bf 12}
  (2007)  069},
\href{http://arxiv.org/abs/0710.4166}{{\tt arXiv:0710.4166 [hep-th]}}.
%%CITATION = 0710.4166;%%.

\bibitem{Vaman:2002ka}
D.~Vaman and H.~L. Verlinde, ``{Bit strings from N=4 gauge theory},''
  \href{http://dx.doi.org/10.1088/1126-6708/2003/11/041}{{\em JHEP} {\bf 0311}
  (2003)  041},
\href{http://arxiv.org/abs/hep-th/0209215}{{\tt arXiv:hep-th/0209215
  [hep-th]}}.
%%CITATION = HEP-TH/0209215;%%.

\bibitem{Dutta:2007ws}
S.~Dutta and R.~Gopakumar, ``{Free Fermions and Thermal AdS/CFT},''
  \href{http://dx.doi.org/10.1088/1126-6708/2008/03/011}{{\em JHEP} {\bf 03}
  (2008)  011},
\href{http://arxiv.org/abs/0711.0133}{{\tt arXiv:0711.0133 [hep-th]}}.
%%CITATION = 0711.0133;%%.

\bibitem{Hamermesh}
M.~Hamermesh, {\em Group theory and its applications to physical problems}.
\newblock Courier Dover Publications, 1989.

\bibitem{Benvenuti:2006qr}
S.~Benvenuti, B.~Feng, A.~Hanany, and Y.-H. He, ``{Counting BPS Operators in
  Gauge Theories: Quivers, Syzygies and Plethystics},''
  \href{http://dx.doi.org/10.1088/1126-6708/2007/11/050}{{\em JHEP} {\bf 0711}
  (2007)  050},
\href{http://arxiv.org/abs/hep-th/0608050}{{\tt arXiv:hep-th/0608050
  [hep-th]}}.
%%CITATION = HEP-TH/0608050;%%.
%
%\bibitem{Feng:2007ur}
B.~Feng, A.~Hanany, and Y.-H. He, ``{Counting gauge invariants: The Plethystic
  program},'' \href{http://dx.doi.org/10.1088/1126-6708/2007/03/090}{{\em JHEP}
  {\bf 0703} (2007)  090},
\href{http://arxiv.org/abs/hep-th/0701063}{{\tt arXiv:hep-th/0701063
  [hep-th]}}.
%%CITATION = HEP-TH/0701063;%%.

\bibitem{Goldstein_2nd_edition}
H.~Goldstein, {\em Classical Mechanics - 2nd Edition}.
\newblock Addison Wesley, 1980.

\bibitem{Eynard:2005wg}
B.~Eynard and A.~P. Ferrer, ``{2-matrix versus complex matrix model, integrals
  over the unitary group as triangular integrals},''
  \href{http://dx.doi.org/10.1007/s00220-006-1541-8}{{\em Commun.Math.Phys.}
  {\bf 264} (2006)  115--144},
\href{http://arxiv.org/abs/hep-th/0502041}{{\tt arXiv:hep-th/0502041
  [hep-th]}}.
%%CITATION = HEP-TH/0502041;%%.

\bibitem{Atick:1988si}
J.~J. Atick and E.~Witten, ``{The Hagedorn Transition and the Number of Degrees
  of Freedom of String Theory},''
\href{http://dx.doi.org/10.1016/0550-3213(88)90151-4}{{\em Nucl.Phys.} {\bf
  B310} (1988)  291--334}.
%%CITATION = NUPHA,B310,291;%%.

\bibitem{Berenstein:2004kk}
D.~Berenstein, ``{A Toy model for the AdS / CFT correspondence},''
  \href{http://dx.doi.org/10.1088/1126-6708/2004/07/018}{{\em JHEP} {\bf 0407}
  (2004)  018},
\href{http://arxiv.org/abs/hep-th/0403110}{{\tt arXiv:hep-th/0403110
  [hep-th]}}.
%%CITATION = HEP-TH/0403110;%%.

\bibitem{Festuccia:2006sa}
G.~Festuccia and H.~Liu, ``{The Arrow of time, black holes, and quantum mixing
  of large N Yang-Mills theories},''
  \href{http://dx.doi.org/10.1088/1126-6708/2007/12/027}{{\em JHEP} {\bf 0712}
  (2007)  027},
\href{http://arxiv.org/abs/hep-th/0611098}{{\tt arXiv:hep-th/0611098
  [hep-th]}}.
%%CITATION = HEP-TH/0611098;%%.
%
%\bibitem{Iizuka:2008hg}
N.~Iizuka and J.~Polchinski, ``{A Matrix Model for Black Hole
  Thermalization},''
  \href{http://dx.doi.org/10.1088/1126-6708/2008/10/028}{{\em JHEP} {\bf 0810}
  (2008)  028},
\href{http://arxiv.org/abs/0801.3657}{{\tt arXiv:0801.3657 [hep-th]}}.
%%CITATION = ARXIV:0801.3657;%%.

\bibitem{Balasubramanian:2005mg}
V.~Balasubramanian, J.~de~Boer, V.~Jejjala, and J.~Simon, ``{The Library of
  Babel: On the origin of gravitational thermodynamics},''
  \href{http://dx.doi.org/10.1088/1126-6708/2005/12/006}{{\em JHEP} {\bf 0512}
  (2005)  006},
\href{http://arxiv.org/abs/hep-th/0508023}{{\tt arXiv:hep-th/0508023
  [hep-th]}}.
%%CITATION = HEP-TH/0508023;%%.

\bibitem{Gutowski:2004ez}
J.~B. Gutowski and H.~S. Reall, ``Supersymmetric {$\ads_5$} black holes,'' {\em
  JHEP} {\bf 02} (2004)  006,
\href{http://arxiv.org/abs/hep-th/0401042}{{\tt hep-th/0401042}}.
%%CITATION = HEP-TH/0401042;%%.
%
%\bibitem{Gutowski:2004yv}
J.~B. Gutowski and H.~S. Reall, ``General supersymmetric {$\ads_5$} black
  holes,'' {\em JHEP} {\bf 04} (2004)  048,
\href{http://arxiv.org/abs/hep-th/0401129}{{\tt hep-th/0401129}}.
%%CITATION = HEP-TH/0401129;%%.

\bibitem{Chang:2013fba}
C.-M. Chang and X.~Yin, ``{1/16 BPS states in $\mathcal N=$ 4 super-Yang-Mills
  theory},'' \href{http://dx.doi.org/10.1103/PhysRevD.88.106005}{{\em
  Phys.Rev.} {\bf D88} (2013) no.~10, 106005},
\href{http://arxiv.org/abs/1305.6314}{{\tt arXiv:1305.6314 [hep-th]}}.
%%CITATION = ARXIV:1305.6314;%%.

\bibitem{Kimura:2009ur}
Y.~Kimura, S.~Ramgoolam, and D.~Turton, ``{Free particles from Brauer algebras
  in complex matrix models},''
  \href{http://dx.doi.org/10.1007/JHEP05(2010)052}{{\em JHEP} {\bf 1005} (2010)
   052},
\href{http://arxiv.org/abs/0911.4408}{{\tt arXiv:0911.4408 [hep-th]}}.
%%CITATION = ARXIV:0911.4408;%%.

\end{thebibliography}

\providecommand{\href}[2]{#2}\begingroup\raggedright\endgroup

\end{document}